\documentclass[10pt]{article}

\usepackage{epsfig}
\usepackage{latexsym}
\usepackage{amssymb}
\usepackage{amsmath}
\usepackage{proof}
\usepackage{pstricks}
\usepackage{times,latex8}
\usepackage{mathrsfs}
\usepackage{stmaryrd}
\usepackage{subfigure}
\usepackage{isolatin1}
\usepackage{url}

\newcommand{\ass}{\mathsf{a}}
\newcommand{\oss}{\mathsf{o}}
\newcommand{\ess}{\mathsf{e}}
\newcommand{\mss}{\mathsf{m}}
\newcommand{\nss}{\mathsf{n}}
\newcommand{\sss}{\mathsf{s}}
\newcommand{\lss}{\mathsf{l}}
\newcommand{\rss}{\mathsf{r}}
\newcommand{\fss}{\mathsf{f}}
\newcommand{\pss}{\mathsf{p}}
\newcommand{\xss}{\mathsf{x}}

\newcommand{\spb}{\;|\;}

\newcommand{\N}{\mathbb{N}}

\newcommand{\ELL}{\ensuremath{\textsf{ELL}}}
\newcommand{\MELL}{\ensuremath{\textsf{MELL}}}
\newcommand{\MELLfull}{\ensuremath{\textsf{MELL}_{\multimap\otimes\forall}}}
\newcommand{\MELLprop}{\ensuremath{\textsf{MELL}_{\multimap\otimes}}}
\newcommand{\MELLimpl}{\ensuremath{\textsf{MELL}_{\multimap}}}

\newcommand{\LLL}{\textsf{LLL}}

\newcommand{\SLL}{\textsf{SLL}}
\newcommand{\QLL}{\textsf{4LL}}
\newcommand{\TLL}{\textsf{TLL}}
\newcommand{\BLL}{\textsf{BLL}}
\newcommand{\IAM}{\textsf{IAM}}
\newcommand{\longrightarrowtriangle}{\ensuremath{\mathop{-\!\!\!\rightarrowtriangle}}}
\newcommand{\dlength}[2]{||#1||_{#2}}
\newcommand{\stepsto}[2]{[#1]_{#2}}
\newcommand{\length}[1]{|#1|}
\newcommand{\tlength}[1]{||#1||}
\newcommand{\plength}[2]{|#1|_{#2}}

\newtheorem{lemma}{Lemma}
\newtheorem{proposition}{Proposition}
\newtheorem{theorem}{Theorem}

\newtheorem{definition}{Definition}

\newenvironment{proof}{\begin{trivlist}
       \item[\hskip \labelsep {\bfseries Proof.}]}{\hfill$\Box$\end{trivlist}}

\newenvironment{lemma*}[1][Lemma]{\begin{trivlist}
       \item[\hskip \labelsep {\bfseries Lemma #1}]\it}{\end{trivlist}}
\newenvironment{proposition*}[1][Proposition]{\begin{trivlist}
       \item[\hskip \labelsep {\bfseries Proposition #1}]\it}{\end{trivlist}}
\newenvironment{theorem*}[1][Theorem]{\begin{trivlist}
       \item[\hskip \labelsep {\bfseries Theorem #1}]\it}{\end{trivlist}}
   
\newcommand{\linear}{\multimap}

\newcommand{\parity}[1]{\mathit{parity}(#1)}

\newcounter{number}
\newenvironment{varitemize}
{
\begin{list}{\labelitemi}
{
\setlength{\itemsep}{0pt}
 \setlength{\topsep}{0pt}
 \setlength{\parsep}{0pt}
 \setlength{\partopsep}{0pt}
 \setlength{\leftmargin}{15pt}
 \setlength{\rightmargin}{0pt}
 \setlength{\itemindent}{0pt}
 \setlength{\labelsep}{5pt}
 \setlength{\labelwidth}{10pt}}}
{
 \end{list} 
}

\newenvironment{varnumlist}
{
\begin{list}{\arabic{number}.}
{\usecounter{number}
 \setlength{\itemsep}{0.0mm}
 \setlength{\topsep}{0.0mm}
 \setlength{\parindent}{0.0mm}
 \setlength{\parskip}{0.0mm}
 \setlength{\parsep}{0.0mm}
 \addtolength{\leftmargin}{-\labelsep}}}
{
 \end{list} 
}

\begin{document}
  \title{Context Semantics, Linear Logic and Computational Complexity\thanks{The 
      author is partially supported by PRIN project
      FOLLIA (2004) and ANR project NOCOST (2005).}} 
  \author{Ugo Dal Lago\\
    {\it Laboratoire d'Informatique de Paris-Nord}\\ 
    {\it Universit\'e Paris 13, France}\\
    \texttt{dallago@lipn.univ-paris13.fr}}
\maketitle
\begin{abstract}
We show that context semantics can be fruitfully applied
to the quantitative analysis of proof normalization
in linear logic. In particular, context semantics lets us
define the \emph{weight} of a proof-net as a measure of
its inherent complexity: it is both an upper bound
to normalization time (modulo a polynomial overhead,
independently on the reduction strategy) and a lower 
bound to the number of steps to
normal form (for certain reduction strategies). Weights are 
then exploited in proving strong soundness theorems 
for various subsystems of linear logic, namely elementary 
linear logic, soft linear logic and light linear logic.
\end{abstract}
\section{Introduction}
Linear logic has always been claimed
to be resource-conscious: structural rules are applicable only when
the involved formulas are modal, i.e.\ in the form $!A$. Indeed, while
(multiplicative) linear logic embeds intuitionistic logic, restricting
the rules governing the exponential operator $!$ leads to characterizations of
interesting complexity classes~\cite{Asperti02tocl,Girard98ic,Lafont04tcs}. On
the other hand, completely forbidding duplication highlights strong relations between
proofs and boolean circuits~\cite{Terui04lics}. These results demonstrate the
relevance of linear logic in implicit computational complexity, where 
the aim is obtaining machine-independent, logic-based characterization of 
complexity classes. Nevertheless, relations between copying and complexity
are not fully understood yet. Is copying the real ``root'' of compleixty?
Can we give a complexity-theoretic interpretation of Girard's 
embedding $A\rightarrow B\equiv !A\linear B$? Bounds on normalization time
for different fragments of linear logic are indeed obtained by ad-hoc
techniques which cannot be easily generalized.

Context semantics~\cite{Gonthier92popl} is a powerful framework for the 
analysis of proof and program dynamics. It can be considered as a model of 
Girard's geometry of interaction~\cite{Girard89lc,Girard88cl} where the
underlying algebra consists of \emph{contexts}. Context semantics and the geometry
of interaction have been used to prove the correctness of optimal reduction
algorithms~\cite{Gonthier92popl} and in the design of sequential and 
parallel interpreters for the lambda calculus~\cite{Mackie95popl,Pinto01tlca}. 
There are evidences that these semantic frameworks are useful in capturing
quantitative as well as qualitative properties of programs. 
The inherent computational difficulty of normalizing a proof has 
indeed direct counterpart in its interpretation. It is well 
known that strongly normalizing proofs are exactly the ones having finitely 
many so-called regular paths in the geometry of interaction~\cite{danos95advances}. 
A class of proof-nets which are not just strongly normalizing but normalizable
in elementary time can still be captured in the geometry of interaction
framework, as suggested by Baillot and Pedicini~\cite{baillot01fi}.
Until recently, it was not known whether this correspondence scales down
to smaller complexity classes, such as the one of polynomial time computable functions.
The usual measure based on the length of regular paths cannot
be used, since there are proof-nets which can be normalized in polynomial time
but whose regular paths have exponential length (as we are
going to show in the following). Context semantics has been 
recently exploited by the author in the quantitative analysis of linear lambda calculi with 
higher-order recursion~\cite{dallago05lics}. Noticeably, context semantics is powerful enough
to induce bounds on the algebraic potential size of terms, a parameter
which itself bounds normalization time (up to a polynomial overhead). 
From existing literature, it is not clear whether similar results can be achieved for 
linear logic, where exponentials take the place of recursion 
in providing the essential expressive power.

In this paper, we show that context semantics reveals precise quantitative information on the 
dynamics of second order multiplicative and exponential linear logic. More
specifically, a weight $W_G$ is assigned to every proof-net $G$ in such a way
that:
\begin{varitemize}
  \item
    Both the number of steps to normal form \emph{and} the size of any reduct 
    of $G$ are bounded by $p(W_G,|G|)$, where $p:\N^2\rightarrow\N$ is a fixed polynomial
    and $|G|$ is the size of $G$.
  \item
    There is a reduction strategy which \emph{realizes} $W_G$, i.e. there is
    a proof-net $H$ such that $G\longrightarrow^{W_G} H$.
\end{varitemize}
In other words, we show that context semantics is somehow ``fully-abstract'' with
respect to the operational theory equating all proofs having the same quantitative
behaviour (modulo a fixed polynomial and considering all reduction strategies).

Moreover, studying $W_G$ is easier than dealing directly with the underlying
syntax. In particular, we here
prove strong soundness theorems (any proof can be reduced in a bounded amount of
time, independently on the underlying reduction strategy) for various subsystems of 
multiplicative linear logic by studying how restricting exponential rules reflect
to $W_G$. These proofs are simpler than similar ones from the 
literature~\cite{Girard98ic,Asperti02tocl,Terui01lics,Lafont04tcs},
which in many cases refer to weak rather than strong soundness.

The weight $W_G$ of a proof-net $G$ will be defined from the context semantics of $G$ 
following two ideas:
\begin{varitemize}
  \item
    The cost of a given box inside $G$ is the number of times it can possibly be copied during
    normalization;
  \item
    The weight of $G$ is the sum of costs of boxes inside $G$, where boxes that are inside
    other boxes are possibly counted more than once.
\end{varitemize}
As a consequence, $W_G$ only takes into account the exponential portion of
$G$ and is null whenever $G$ does not contain any instance of the exponential rules.

We are going to define context semantics in a style which is very reminiscent of
the one used by Danos and Regnier when defining their interaction abstract
machine (\IAM, see~\cite{Danos99tcs}). There are, however, some additional rules
that makes the underlying machine not strictly bideterministic.
As we will detail in the rest of the paper, the added transition rules 
are essential to capture the quantitative behaviour of proofs under every
possible reduction strategy. 

The rest of this paper is organized as follows. In Section~\ref{sect:syntax}, we will
define linear logic as a sequent calculus and as a system of proof-nets. 
In Section~\ref{sect:semantics}, context semantics is defined and some 
examples of proof-nets are presented, together with their interpretation. Section~\ref{sect:complexity}
is devoted to relationships between context semantics and computational complexity and
presents the two main results. Section~\ref{sect:subsystems} describe how context
semantics can be useful in studying subsystems of linear logic, namely elementary
linear logic, soft linear logic and light linear logic. 
This is the full version of a recently appeared extended-abstract~\cite{Dallago06lics}.

\section{Syntax}\label{sect:syntax}
We here introduce multiplicative linear logic as 
a sequent calculus. Then, we will
show how a proof-net can be associated to any
sequent-calculus proof. The results described in
the rest of this paper are formulated in terms of
proof-nets.\par
The language of \emph{formulae} is
defined by the following productions:
$$
A::=\alpha\spb A\linear A\spb A\otimes A\spb !A\spb\forall\alpha.A
$$
where $\alpha$ ranges over a countable set of \emph{atoms}.
The rules in Figure~\ref{figure:TYPEASSIGN}
define a sequent calculus for (intuitionistic) multiplicative and
exponential linear logic (with second order).
\begin{figure*}
\begin{center}
\input{typeassignment}
\caption{A sequent calculus for \MELL}
\label{figure:TYPEASSIGN}
\end{center}
\end{figure*}
We shall use \MELL\ or \MELLfull\ as a shorthand for this
system. In this way, we are able to
easily identify interesting fragments, such as
the propositional fragment \MELLprop\ or the
implicative fragment \MELLimpl. Observe the
Girard's translation $A\rightarrow B\equiv !A\linear B$ 
enforces the following embeddings:
\begin{varitemize}
  \item
    Simply-typed lambda calculus into
    \MELLimpl.
  \item
    Intuitionistic propositional logic into
    \MELLprop.
  \item
    Intuitionistic second-order logic into
    \MELLfull.
\end{varitemize}

Proof-nets~\cite{Girard95pn} are graph-like representations
for proofs. We here adopt a system of intuitionistic
proof-nets; in other words, we do not map derivations in 
\MELLfull\ to usual, classical, proof-nets.\par
\begin{figure}
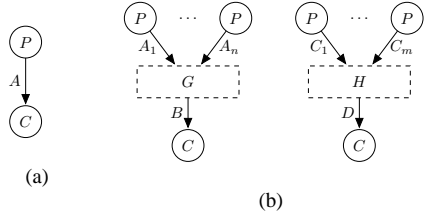

\begin{center}
\subfigure[]{
  \begin{minipage}[c]{13.2pt}
    \centering\scalebox{0.6}{\epsfbox{figure.8}}
  \end{minipage}
}
\hspace{20pt}
\subfigure[]{
  \begin{minipage}[c]{49.8pt}
    \centering\scalebox{0.6}{\epsfbox{figure.2}}
  \end{minipage}
  \hspace{10pt}
  \begin{minipage}[c]{49.8pt}
    \centering\scalebox{0.6}{\epsfbox{figure.58}}
  \end{minipage}
}
\caption{Base cases.}
\label{fig:baseinduction}
\end{center}
\end{figure}
\begin{figure*}
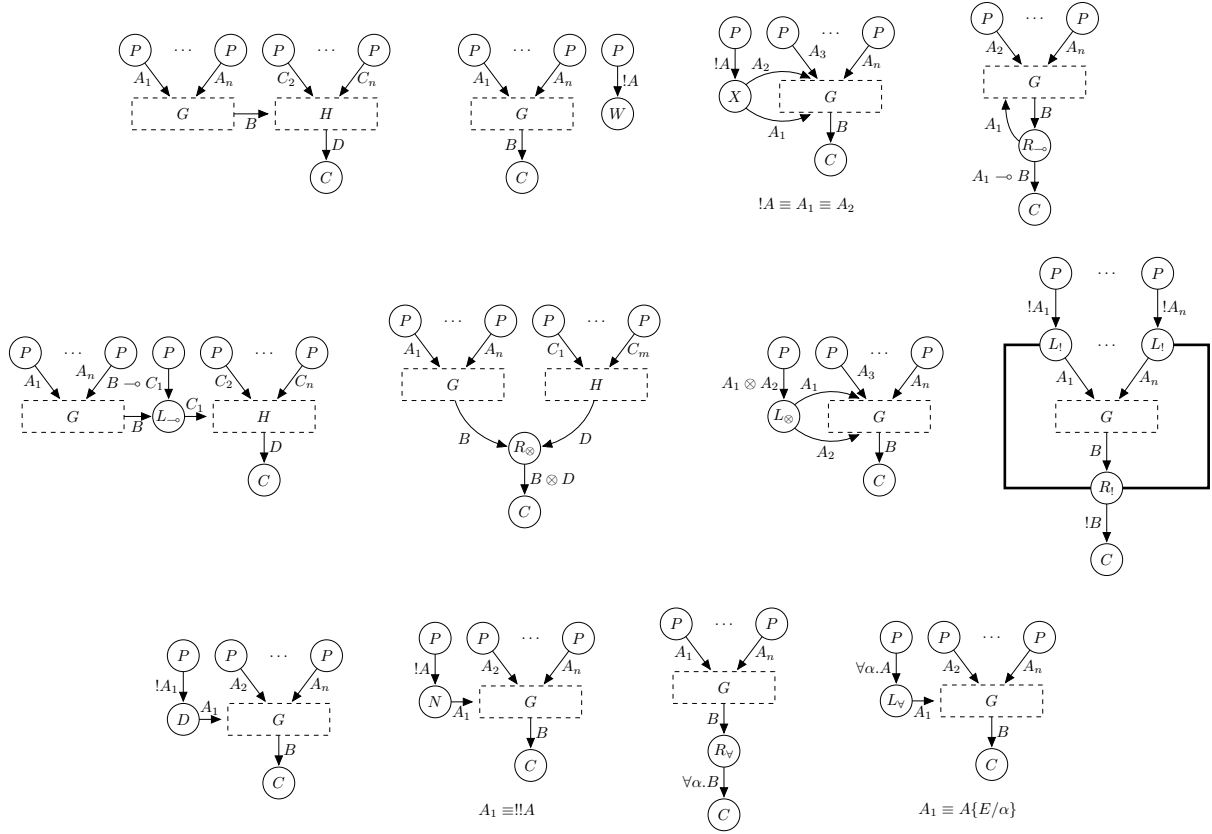
 
\begin{center}
  \begin{minipage}[c]{103.2pt}
    \centering\scalebox{0.6}{\epsfbox{figure.59}}
  \end{minipage} \hspace{20pt}
  \begin{minipage}[c]{69pt}
    \centering\scalebox{0.6}{\epsfbox{figure.10}}
  \end{minipage} \hspace{20pt}
  \begin{minipage}[c]{75pt}
    \centering\scalebox{0.6}{\epsfbox{figure.25}}
  \end{minipage} \hspace{20pt}
  \begin{minipage}[c]{49.2pt}
    \centering\scalebox{0.6}{\epsfbox{figure.3}}
  \end{minipage}\\\vspace{10pt}
  \begin{minipage}[c]{121.2pt}
    \centering\scalebox{0.6}{\epsfbox{figure.4}}
  \end{minipage}\hspace{20pt}
  \begin{minipage}[c]{103.2pt}
    \centering\scalebox{0.6}{\epsfbox{figure.17}}
  \end{minipage}\hspace{20pt}
  \begin{minipage}[c]{84.6pt}
    \centering\scalebox{0.6}{\epsfbox{figure.16}}
  \end{minipage}\hspace{20pt}
  \begin{minipage}[c]{78pt}
    \centering\scalebox{0.6}{\epsfbox{figure.7}}
  \end{minipage}\\\vspace{10pt}
  \begin{minipage}[c]{71.6pt}
    \centering\scalebox{0.6}{\epsfbox{figure.12}}
  \end{minipage} \hspace{20pt}
  \begin{minipage}[c]{71.6pt}
    \centering\scalebox{0.6}{\epsfbox{figure.13}}
  \end{minipage}\hspace{20pt}
  \begin{minipage}[c]{49.2pt}
    \centering\scalebox{0.6}{\epsfbox{figure.50}}
  \end{minipage} \hspace{20pt}
  \begin{minipage}[c]{76.2pt}
    \centering\scalebox{0.6}{\epsfbox{figure.51}}
  \end{minipage} \hspace{20pt}
\end{center}
\caption{Inductive cases}
\label{fig:induction}
\end{figure*}

Let $\mathscr{L}$ be the set 
$$
\{R_\linear,L_\linear,R_\otimes,L_\otimes,R_\forall,L_\forall,R_!,L_!,W,X,D,N,P,C\}
$$ 
A {\it proof-net} is a graph-like structure $G$. It
can be defined inductively as follows: a proof-net is either the
graph in Figure~\ref{fig:baseinduction}(a) or one of those in Figure~\ref{fig:induction} where 
$G,H$ are themselves proof-nets as in Figure~\ref{fig:baseinduction}(b).
If $G$ is a proof-net, then $V_G$ denotes the set of vertices
of $G$, $E_G$ denotes the set of direct edges of $G$,
$\alpha_G$ is a labelling functions mapping every vertex in $V_G$ to an 
element of $\mathscr{L}$ and $\beta_G$ maps every edge in $E_G$ to a formula.
We do not need to explicitly denote axioms and cuts by vertices in $V_G$.\par
Note that each of the rules in figures~\ref{fig:baseinduction}(a) 
and~\ref{fig:induction} closely corresponds to a rule 
in the sequent calculus. Given a sequent calculus proof $\pi$, 
a proof-net $G_\pi$ corresponding to $\pi$ can be built.
We should always be able to distinguish edges which are incident to a vertex
$v$. In particular, we assume the existence of an order between them, which
corresponds to the clockwise order in the graphical representation of $v$.
Nodes labelled with $C$ (respectively, $P$) mark
the conclusion (respectively, the premises) of
the proof-net.
Notice that the rule corresponding to
$P_!$ (see Figure~\ref{fig:induction}) allows 
seeing interaction graphs
as nested structures, where nodes labelled
with $R_!$ and $L_!$ delimit a \emph{box}. If $e\in E_G$,
$\theta_G(e)$ denotes the vertex labelled with $R_!$ 
delimiting the box containing $e$ 
(if such a box exists, otherwise $\theta_G(e)$ is undefined). 
If $v\in V_G$, $\theta_G(v)$ has the same meaning.
If $v$ is a vertex with $\alpha_G(v)=R_!$, then 
$\rho_G(v)$ denotes 
the edge departing from $v$ and going outside the box. 
Expressions $\sigma_G(e)$
and $\sigma_G(v)$ are shorthand for $\rho_G(\theta_G(e))$
and $\rho_G(\theta_G(v))$, respectively. 

If $e=(u,v)\in E_G$,
and $\alpha_G(u)=R_!$, then $e$ is said to be a \emph{box-edge}.
$B_G$ is the set of all box-edges of $G$. Given a box-edge
$e$, $P_G(e)$ is the number of premises of the box. $I_G$ is
the set of all vertices $v\in V_G$ with 
$\alpha_G(v)\notin\{R_!,L_!\}$. If $v\in V_G$, then $\partial(v)$ is
the so-called box-depth of $v$, i.e. the number
of boxes where $v$ is included; similarly, $\partial(e)$
is the box-depth of $e\in E_G$, while $\partial(G)$
is the box-depth of the whole proof-net $G$. The size
$|G|$ of a proof-net $G$ is simply $|V_G|$. 

\begin{figure}[!h]
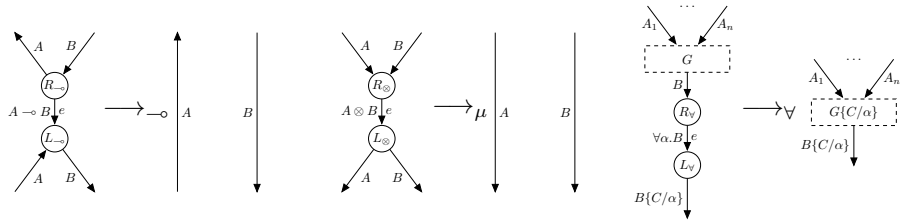

\begin{center}
  \begin{minipage}[c]{34pt}
    \centering\scalebox{0.5}{\epsfbox{figure.14}}
  \end{minipage}
  \begin{minipage}[c]{20pt}
    \centering $\longrightarrow_\linear$
  \end{minipage}
  \begin{minipage}[c]{41.5pt}
    \centering\scalebox{0.5}{\epsfbox{figure.15}}
  \end{minipage}
  \hspace{20pt}
  \begin{minipage}[c]{33pt}
    \centering\scalebox{0.5}{\epsfbox{figure.54}}
  \end{minipage}
  \begin{minipage}[c]{20pt}
    \centering $\longrightarrow_\mu$
  \end{minipage}
  \begin{minipage}[c]{29pt}
    \centering\scalebox{0.5}{\epsfbox{figure.55}}
  \end{minipage}
  \hspace{20pt}
  \begin{minipage}[c]{38pt}
    \centering\scalebox{0.5}{\epsfbox{figure.56}}
  \end{minipage}
  \begin{minipage}[c]{20pt}
    \centering $\longrightarrow_\forall$
  \end{minipage}
  \begin{minipage}[c]{38pt}
    \centering\scalebox{0.5}{\epsfbox{figure.57}}
  \end{minipage}
\caption{Linear graph rewriting rules.}
\label{fig:linearrules}
\end{center}
\end{figure}

\begin{figure}[!h]
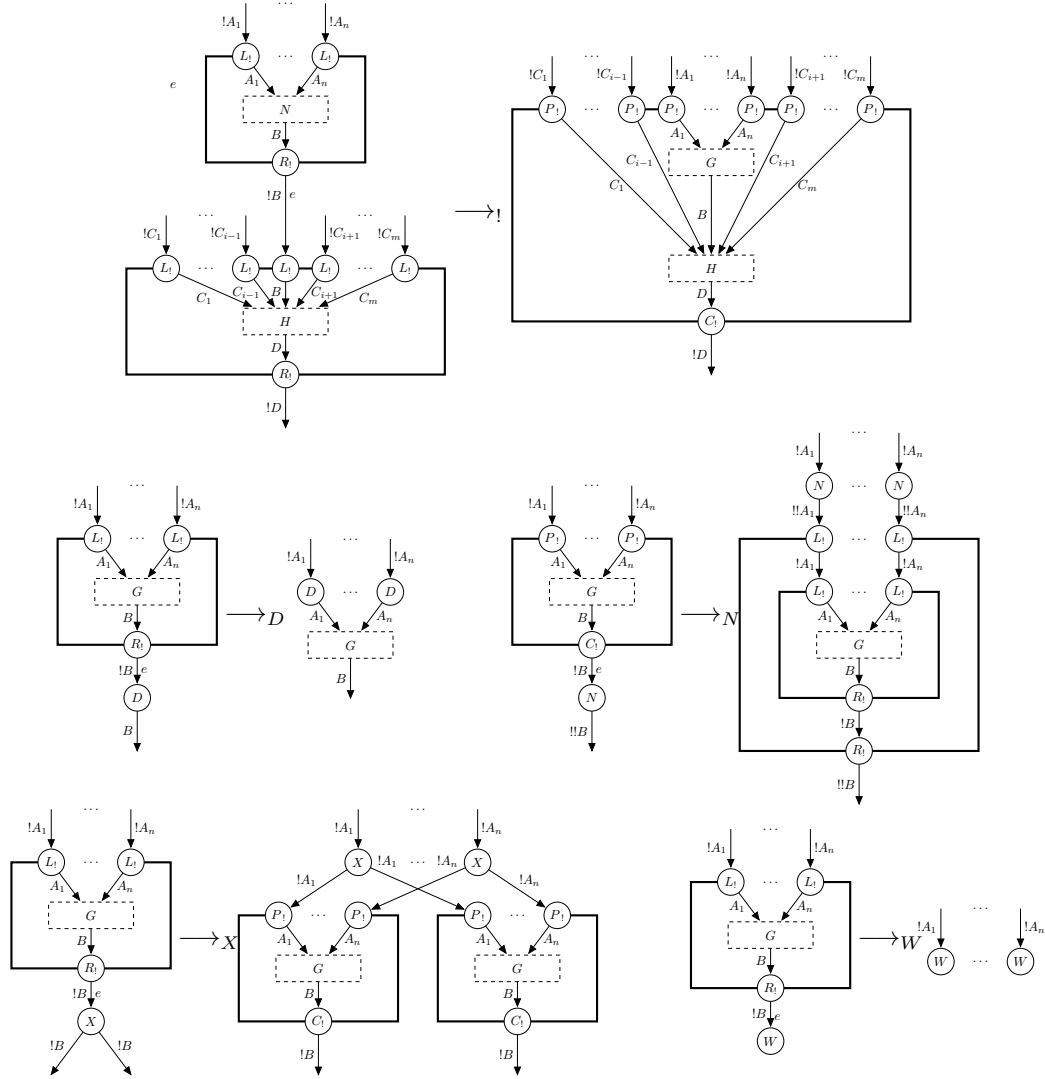

\begin{center}
  \begin{minipage}[c]{121pt}
    \centering\scalebox{0.5}{\epsfbox{figure.23}}
  \end{minipage}
  \begin{minipage}[c]{20pt}
    \centering $\longrightarrow_!$
  \end{minipage}
  \begin{minipage}[c]{151pt}
    \centering\scalebox{0.5}{\epsfbox{figure.24}}
  \end{minipage}\\
  \begin{minipage}[c]{61pt}
    \centering\scalebox{0.5}{\epsfbox{figure.40}}
  \end{minipage}
  \begin{minipage}[c]{20pt}
    \centering $\longrightarrow_D$
  \end{minipage}
  \begin{minipage}[c]{51pt}
    \centering\scalebox{0.5}{\epsfbox{figure.41}}
  \end{minipage}
  \hspace{30pt}
  \begin{minipage}[c]{61pt}
    \centering\scalebox{0.5}{\epsfbox{figure.42}}
  \end{minipage}
  \begin{minipage}[c]{20pt}
    \centering $\longrightarrow_N$
  \end{minipage}
  \begin{minipage}[c]{91pt}
    \centering\scalebox{0.5}{\epsfbox{figure.43}}
  \end{minipage}\\
  \begin{minipage}[c]{61pt}
    \centering\scalebox{0.5}{\epsfbox{figure.44}}
  \end{minipage}
  \begin{minipage}[c]{20pt}
    \centering $\longrightarrow_X$
  \end{minipage}
  \begin{minipage}[c]{136pt}
    \centering\scalebox{0.5}{\epsfbox{figure.45}}
  \end{minipage}
  \hspace{30pt}
  \begin{minipage}[c]{61pt}
    \centering\scalebox{0.5}{\epsfbox{figure.21}}
  \end{minipage}
  \begin{minipage}[c]{20pt}
    \centering $\longrightarrow_W$
  \end{minipage}
  \begin{minipage}[c]{41pt}
    \centering\scalebox{0.5}{\epsfbox{figure.22}}
  \end{minipage}
\caption{Exponential graph rewriting rules.}
\label{fig:exporules}
\end{center}
\end{figure}

Cut elimination is performed by graph rewriting in
proof-nets. There are eight different rewriting rules
$\longrightarrow_S$, where 
$S\in\mathscr{C}=\{\linear,\otimes,\forall,!,X,D,N,W\}$.
We distinguish three linear rewriting rules (see
figure~\ref{fig:linearrules}) and five exponential
rewriting rules (see figure~\ref{fig:exporules}).
If $\mathscr{Q}\subseteq\mathscr{C}$, then $\longrightarrow_\mathscr{Q}$
is the union of $\longrightarrow_S$ over $S\in\mathscr{Q}$. The relation $\longrightarrow$
is simply $\longrightarrow_\mathscr{C}$. The notion of a normal form proof-net 
is the usual one. 
A \emph{cut edge} is the edge linking two nodes interacting in a 
cut-elimination step. In figures~\ref{fig:linearrules}
and~\ref{fig:exporules}, $e$ is always a cut edge. If $S\in\mathscr{C}$,
an edge linking two nodes that interact in $\longrightarrow_S$
is called an $S$-cut. 

Given a proof-net $G$, the expression
$\dlength{G}{\rightarrow}$ denotes the natural number
$$
\max_{G\longrightarrow^* H}|H|.
$$
The expression $\stepsto{G}{\rightarrow}$ denotes the natural
number
$$
\max_{G\longrightarrow^n H}n.
$$
These are well-defined concepts, since the calculus is strongly
normalizing.\par
The relation $\Longrightarrow$ is a restriction on $\longrightarrow$
defined as follows: $G\Longrightarrow H$ iff $G\longrightarrow_S H$
where $S=W$ only in case any cut in $G$ is a $W$-cut. This is
a reduction strategy, i.e. $G\Longrightarrow^* H$ whenever $H$
is the normal form of $G$. Indeed, firing a $W$-cut can only introduce
other $W$-cuts. The expressions $\dlength{G}{\Rightarrow}$ 
and $\stepsto{G}{\Rightarrow}$ are
defined in the obvious way, similarly to $\dlength{G}{\rightarrow}$
and $\stepsto{G}{\rightarrow}$.
Studying $\Longrightarrow$ is easier than studying $\longrightarrow$.
From a complexity point of view, this is not problematic, since
results about $\Longrightarrow$ can be easily transferred to $\longrightarrow$:
\begin{lemma}[Standardization]\label{lemma:standardization}
For every proof-net $G$, both $\stepsto{G}{\Rightarrow}=\stepsto{G}{\rightarrow}$ 
and $\dlength{G}{\Rightarrow}=\dlength{G}{\rightarrow}$.
\end{lemma}
\begin{proof}
Whenever $G\longrightarrow_W H\longrightarrow_S J$ and $S\neq W$,
there are $K$ and $n\in\{1,2\}$ such that $G\longrightarrow_S K\longrightarrow_W^n J$, 
because the two cut-elimination steps do not overlap with each other.
As a consequence, for any sequence $M_1\longrightarrow\ldots\longrightarrow M_n$ there
is another sequence $L_1\Longrightarrow\ldots\Longrightarrow L_m$ such
that $L_1=M_1$, $L_m=M_n$ and $m\geq n$. This proves the first claim.
Now, observe that for any $1\leq i\leq n$ there is $j$ such
that $|L_j|\geq |M_i|$: at any step a proof-net, $H$,
disappears from the sequence being replaced by another one, $K$,
but clearly $|G|\geq |H|$. This concludes the proof.
\end{proof}
In the following, we will prove combinatorial properties of $\Longrightarrow$ that, 
by lemma~\ref{lemma:standardization}, can be easily transferred to $\longrightarrow$. 
Consider the following further conditions on 
$\Longrightarrow$:
\begin{varnumlist}
  \item
    For every $n\in\N$, a cut at level $n+1$ is fired only when
    any cut at levels from $1$ to $n$ is a $W$-cut.
  \item
    For every $n\in\N$, a $!$-cut at level $n$ is fired only when
    any cut at level $n$ is either a $W$-cut or a $!$-cut
\end{varnumlist}
These two conditions induce another relation $\longrightarrowtriangle$,
which is itself a reduction strategy: note that firing
a cut at level $n$ does not introduce cuts at levels strictly smaller than
$n$, while firing a $!$-cut at level $n$ only introduces cuts at level
$n+1$. As a consequence, $\longrightarrowtriangle$ can be considered as
a ``level-by-level'' strategy~\cite{Asperti02tocl,Terui01lics}.

\section{Context Semantics}\label{sect:semantics}
In this section, the context semantics of 
proof-nets is studied. The context semantics of a proof-net $G$
allows to isolate certain paths among those in $G$, called
\emph{persistent} in the literature; studying the length and numerosity of
persistent paths for a proof-net $G$ helps inferring useful quantitative 
properties of $G$.

The first preliminary concept is that of an exponential
signature. Exponential signatures are trees
whose nodes are labelled with symbols $\ess,\rss,\lss,\pss,\nss$.
They serve as contexts while constructing a path in a proof-net,
similarly to what \emph{context marks} do in Gonthier, Abadi and L\'evy's
framework~\cite{Gonthier92lics}. Label $\pss$ has a special role and helps
capturing the tricky combinatorial behavior of rule $N_!$ (see Figure~\ref{figure:TYPEASSIGN}).
For similar reasons, a binary relation $\sqsubseteq$ on exponential signatures is
needed.
\begin{definition}
\begin{varitemize}
\item The language $\mathscr{E}$ of \emph{exponential signatures}
  is defined by induction from the following sets of productions:
  $$
  t,u,v,w::=\ess\spb\rss(t)\spb\lss(t)\spb\pss(t)\spb\nss(t,t). 
  $$
\item
  A \emph{standard} exponential signature is one that does not
  contain the constructor $\pss$. An exponential signature $t$ is
  \emph{quasi-standard} iff for every subtree $\nss(u,v)$ of
  $t$, the exponential signature $v$ is standard.
\item 
  The binary relation $\sqsubseteq$ on $\mathscr{E}$ is
  defined as follows:
  \begin{eqnarray*}
    \ess&\sqsubseteq&\ess;\\
    \rss(t)&\sqsubseteq&\rss(u)\Leftrightarrow t\sqsubseteq u;\\
    \lss(t)&\sqsubseteq&\lss(u)\Leftrightarrow t\sqsubseteq u;\\
    \pss(t)&\sqsubseteq&\pss(u)\Leftrightarrow t\sqsubseteq u;\\
    \pss(t)&\sqsubseteq&\nss(u,v)\Leftrightarrow t\sqsubseteq v;\\
    \nss(t,u)&\sqsubseteq&\nss(v,w)\Leftrightarrow t\sqsubseteq v\mbox{ and }u=w.
  \end{eqnarray*}
  If $u\sqsubseteq t$ then $u$ is a \emph{simplification}
  of $t$.
\item 
  A \emph{stack element} is either an exponential signature or one
  of the following characters: $\ass,\oss,\sss,\fss,\xss$. $\mathscr{S}$ is
  the set of stack elements. $\mathscr{S}$ is ranged over by $s,r$.
\item
  A \emph{polarity} is either $+$ or $-$. $\mathscr{B}$ is
  the set of polarities. The following notation is useful:
  $+\!\downarrow$ is $-$, while $-\!\downarrow$ is $+$.
  If $n$ is a natural number $\parity{n}=+$ if $n$ is even,
  while $\parity{n}=0$ if $n$ is odd.
\item
  If $U\in\mathscr{S}^*$, then $\tlength{U}$ denotes the number of exponential signatures
  in $U$. If $s\in\{\ass,\oss,\sss,\fss,\xss\}$, then $\plength{U}{s}$ is the
  number of occurrences of $s$ in $U$.
\end{varitemize}
\end{definition}
Please observe that if $t$ is standard and $t\sqsubseteq u$, then $t=u$.
Moreover, if $t$ is quasi-standard and $u\sqsubseteq u$, then $u$ is
quasi-standard, too. The structure $(\mathscr{E},\sqsubseteq)$ is a partial order:
\begin{lemma}
The relation $\sqsubseteq$ is reflexive, transitive and antisymmetric.
\end{lemma}
\begin{proof}
The fact $t\sqsubseteq t$ can be proved by an induction on $t$. Similarly,
if $t\sqsubseteq u$ and $u\sqsubseteq t$, then $t=u$ by induction on $t$.
Finally, if $t\sqsubseteq u$ and $u\sqsubseteq v$, then $t\sqsubseteq v$
by induction on $t$.
\end{proof}
We are finally ready to define the context semantics for a proof-net $G$.
Given a proof-net $G$, the set of \emph{contexts} for $G$ is
$$
C_G=E_G\times \mathscr{E}^*\times\mathscr{S}^+\times\mathscr{B}.
$$
Vertices of $G$ with labels $R_\linear,L_\linear,R_\otimes,L_\otimes,R_\forall,L_\forall,$ $R_!,L_!,
X,D,N$ induce rewriting rules on $C_G$. These rules are detailed 
in Table~\ref{table:rwlinear} and Table~\ref{table:rwexpo}. For any
such rule
$$
(e,U,V,b)\longmapsto_G(g,W,Z,c),
$$ 
the \emph{dual} rule
$$
(g,W,Z,c\!\downarrow)\longmapsto_G(e,U,V,b\!\downarrow).
$$ 
holds as well. In other words, relation $\longmapsto_G$ is the smallest binary 
relation on $C_G$ including every instance of rules in Table~\ref{table:rwlinear} and Table~\ref{table:rwexpo},
together with every instance of their duals.

\begin{table*}
\caption{Rewrite Rules for Vertices $R_\linear$, $L_\linear$,$R_\otimes$, $L_\otimes$, $R_\forall$, $L_\forall$}\label{table:rwlinear}
\vspace{10pt}
\begin{center}
\begin{tabular}{|c|c|}\hline\hline
\begin{minipage}[c]{3cm}
\vspace{.3cm}
\centering\scalebox{0.60}{\epsfbox{figure.26}}\\
\vspace{.3cm}
\end{minipage}
&
\begin{minipage}[c]{9cm}
\begin{eqnarray*}
(e,U,V,-)&\longmapsto_G&(h,U,V\cdot\ass,+)\\
(g,U,V,+)&\longmapsto_G&(h,U,V\cdot\oss,+)\\
\end{eqnarray*}
\end{minipage}
\\ \hline
\begin{minipage}[c]{3cm}
\vspace{.3cm}
\centering\scalebox{0.60}{\epsfbox{figure.27}}\\
\vspace{.3cm}
\end{minipage}
 &
\begin{minipage}[c]{9cm}
\begin{eqnarray*}
(e,U,V\cdot\ass,+)&\longmapsto_G&(g,U,V,-)\\
(e,U,V\cdot\oss,+)&\longmapsto_G&(h,U,V,+)\\
\end{eqnarray*}
\end{minipage}
\\ \hline
\begin{minipage}[c]{3cm}
\vspace{.3cm}
\centering\scalebox{0.60}{\epsfbox{figure.38}}\\
\vspace{.3cm}
\end{minipage}
 &
\begin{minipage}[c]{9cm}
\begin{eqnarray*}
(e,U,V,+)&\longmapsto_G&(h,U,V\cdot\fss,+)\\
(g,U,V,+)&\longmapsto_G&(h,U,V\cdot\xss,+)\\
\end{eqnarray*}
\end{minipage}
\\ \hline
\begin{minipage}[c]{3cm}
\vspace{.3cm}
\centering\scalebox{0.60}{\epsfbox{figure.39}}\\
\vspace{.3cm}
\end{minipage}
 &
\begin{minipage}[c]{9cm}
\begin{eqnarray*}
(h,U,V\cdot\fss,+)&\longmapsto_G&(e,U,V,+)\\
(h,U,V\cdot\xss,+)&\longmapsto_G&(g,U,V,+)\\
\end{eqnarray*}
\end{minipage}
\\ \hline
\begin{minipage}[c]{3cm}
\vspace{.3cm}
\centering\scalebox{0.60}{\epsfbox{figure.36}}\\
\vspace{.3cm}
\end{minipage}
 &
\begin{minipage}[c]{9cm}
\begin{eqnarray*}
(e,U,V,+)&\longmapsto_G&(g,U,V\cdot\sss,+)\\
\end{eqnarray*}
\end{minipage}
\\ \hline
\begin{minipage}[c]{3cm}
\vspace{.3cm}
\centering\scalebox{0.60}{\epsfbox{figure.37}}\\
\vspace{.3cm}
\end{minipage}
 &
\begin{minipage}[c]{9cm}
\begin{eqnarray*}
(e,U,V\cdot\sss,+)&\longmapsto_G&(g,U,V,+)\\
\end{eqnarray*}
\end{minipage}
\\ \hline\hline
\end{tabular}
\end{center}
\end{table*}
\begin{table*}
\begin{center}
\caption{Rewrite Rules for Vertices $X$, $D$, $N$, $L_!$ and $R_!$.}\label{table:rwexpo}
\vspace{10pt}
\begin{tabular}{|c|c|}\hline\hline
\begin{minipage}[c]{3cm}
\vspace{.3cm}
\centering\scalebox{0.60}{\epsfbox{figure.30}}\\
\vspace{.3cm}
\end{minipage}
 &
\begin{minipage}[c]{9cm}
\begin{eqnarray*}
(h,U,V\cdot \lss(t),+)&\longmapsto_G&(e,U,V\cdot t,+)\\
(h,U,V\cdot \rss(t),+)&\longmapsto_G&(g,U,V\cdot t,+)\\
\end{eqnarray*}
\end{minipage}
\\ \hline
\begin{minipage}[c]{3cm}
\vspace{.3cm}
\centering\scalebox{0.60}{\epsfbox{figure.35}}\\
\vspace{.3cm}
\end{minipage}
 &
\begin{minipage}[c]{9cm}
\begin{eqnarray*}
(e,U,V\cdot\ess,+)&\longmapsto_G&(g,U,V,+)\\
\end{eqnarray*}
\end{minipage}
\\ \hline
\begin{minipage}[c]{3cm}
\vspace{.3cm}
\centering\scalebox{0.60}{\epsfbox{figure.31}}\\
\vspace{.3cm}
\end{minipage}
 &
\begin{minipage}[c]{9cm}
\begin{eqnarray*}
(e,U,V\cdot\nss(t,u),+)&\longmapsto_G&(g,U,V\cdot t\cdot u,+)\\
(e,U,\pss(t),+)&\longmapsto_G&(g,U,t,+)\\
\end{eqnarray*}
\end{minipage}
\\ \hline
\begin{minipage}[c]{3cm}
\vspace{.3cm}
\centering\scalebox{0.60}{\epsfbox{figure.34}}\\
\vspace{.3cm}
\end{minipage}
 &
\begin{minipage}[c]{9cm}
\begin{eqnarray*}
(e,U,V\cdot t,+)&\longmapsto_G&(g,U\cdot t,V,+)\\
(l,U\cdot t,V,+)&\longmapsto_G&(h,U,V\cdot t,+)\\
(e,U,t,+)&\longmapsto_G&(h,U,t,+)\\
\end{eqnarray*}
\end{minipage}
\\ \hline\hline
\end{tabular}
\end{center}
\end{table*}
The role of the four components of a 
context can be intuitively explained as follows:
\begin{varitemize}
\item
  The first component is an edge in the proof-net $G$.
  As a consequence, from every sequence 
  $C_1\longmapsto_G C_2\longmapsto_G\ldots\longmapsto_G C_n$
  we can extract a sequence $e_1,e_2,\ldots,e_n$ of edges.
  Rewriting rules in tables~\ref{table:rwlinear}
  and~\ref{table:rwexpo} enforce this sequence to be
  a path in $G$, i.e. $e_i$ has a vertex in common with 
  $e_{i+1}$. The only exception is caused by the 
  last rule induced by boxes (see Table~\ref{table:rwexpo}): 
  in that case $e$ and $h$ do not share any vertex, but
  the two vertices $v$ and $w$ (which are adjacent
  to $e$ and $h$, respectively) are part of the same box.
\item
  The second component is a (possibly empty) sequence
  of exponential signatures which keeps track of which copies
  of boxes we are currently traveling into. More
  specifically, if $e$ and $U$ are the first and second
  components of a context, then the $\partial(e)$
  rightmost exponential signatures in $U$ correspond to
  copies of the $\partial(e)$ boxes where $e$ is
  contained. Although the definition of a context does
  not prescribe this correspondence ($U$ can be empty
  even if $\partial(e)$ is strictly positive), it is 
  preserved by rewriting.
\item
  The third component is a nonempty sequence of stack elements.
  It keeps track of the history of previously visited edges.
  In this way, the fundamental property called \emph{path-persistence}
  is enforced: any path induced by the context semantics
  is preserved by normalization~\cite{Gonthier92lics}. This
  property is fundamental for proving the correctness of
  optimal reduction algorithms~\cite{Gonthier92popl}, but it
  is not directly exploited in this paper. Notice that
  exponential signatures can float from the second component
  to the third component and vice versa (see the rules 
  induced by vertices $R_!$ and $L_!$).
\item
  The only purpose of the last component is forcing rewriting to
  be (almost) deterministic: for every $C$ there is
  at most one context $D$ such that $C\longmapsto_G D$, except when
  $C=(e,U,t,-)$ and $e\in B_G$. In fact, $C\longmapsto_G (g_i,U,t,-)$ for every $i$,
  where $g_1,\ldots,g_n$ are the premises of the box whose conclusion
  is $e$.
\end{varitemize}
The way we have defined context semantics, namely by a set of contexts
endowed with a rewrite relation, is fairly 
standard~\cite{Gonthier92popl,Gonthier92lics}. 
In particular, our definition owes much to Danos and Regnier's 
Interaction Abstract Machine (\IAM, see~\cite{Danos99tcs}).
Both our machinery and the \IAM\ are reversible, but while
\IAM\ can be considered as a bideterministic automaton, 
our context semantics cannot, due to the last rule induced by boxes.
Noticeably, a fragment of \MELL\ called light linear logic does
enforce strong determinacy, as we will detail in Section~\ref{sect:subsystems}.
A property that holds for \IAM\ as well as for our context semantics
is reversibility: If $(e,U,V,b)\longmapsto_G (g,W,Z,c)$,
then $(g,W,Z,c\!\downarrow)\longmapsto_G (e,U,V,b\!\downarrow)$.

Although context semantics can be defined on sharing graphs as well, 
proof-nets have been considered here. Indeed, sharing graphs are more problematic 
from a complexity viewpoint, since a computationally expensive read-back 
procedure is necessary in order to retrieve the proof (or term) corresponding 
to a sharing graph in normal form (see~\cite{Asperti98book}).

Observe that the semantic framework we have just introduced is \emph{not}
a model of geometry of interaction as described by Girard~\cite{Girard89lc}. In
particular, jumps between distinct conclusions of a box are not permitted in
geometry of interaction, which is completely local in this sense. Moreover, algebraic 
equations induced by rule $N$ are here slightly different. As we are going to see,
this mismatch is somehow necessary in order to capture the combinatorial behavior of
proofs independently on the underlying reduction strategy.

\subsection{Motivating Examples}
We now define some proof-nets together with
observations about how context-semantics reflects
the complexity of normalization.\par
The first example (due to Danos and Regnier) is somehow 
discouraging: a family of 
proof-nets which normalize in polynomial time having
paths of exponential lengths. For every positive natural number 
$n$ and for every formula $A$, a proof-net $G_n(A)$ can be defined.
We go by induction on $n$:
\begin{varitemize}
  \item
    The proof-net $G_1(A)$ is the following:
    \begin{center}
      \scalebox{0.5}
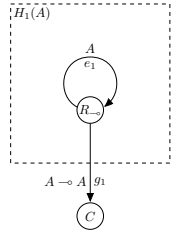
{\epsfbox{figure.79}}
    \end{center}
    Notice we have implicitly defined a sub-graph
    $H_1(A)$ of $G_1(A)$.
  \item
    If $n>1$, then $G_n(A)$ is the following proof-net:
    \begin{center}
      \scalebox{0.5}
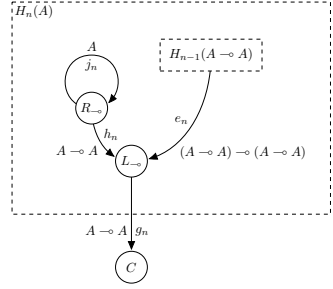
{\epsfbox{figure.80}}
    \end{center}
    Notice we have implicitly defined a sub-graph
    $H_n(A)$ of $G_n(A)$.
\end{varitemize}
Although the size of $G_n(A)$ is $2n$ and, most important,
$G_n(A)$ normalizes in $n-1$ steps to $G_1(A)$, we can easily prove the
following, surprising, fact: for every $n$, for every $A$ and for every 
$V\in\mathscr{S}^*$,
\begin{eqnarray*}
(g_n,\varepsilon,V\cdot\ass,-)&\longmapsto_{G_n(A)}^{f(n)}&(g_n,\varepsilon,V\cdot\oss,+)\\
(g_n,\varepsilon,V\cdot\oss,-)&\longmapsto_{G_n(A)}^{f(n)}&(g_n,\varepsilon,V\cdot\ass,+)
\end{eqnarray*}
where $f(n)=O(2^n)$. Indeed, let $f(n)=8\cdot 2^{n-1}-6$ for every $n\geq 1$ 
and proceed by an easy induction on $n$:
\begin{varitemize}
\item
  If $n=1$, then
  \begin{eqnarray*}
    (g_1,\varepsilon,V\cdot\ass,-)&\longmapsto_{G_1(A)}&(e_1,\varepsilon,V,+)\longmapsto_{G_1(A)}(g_1,\varepsilon,V\cdot\oss,+)\\
    (g_1,\varepsilon,V\cdot\oss,-)&\longmapsto_{G_1(A)}&(e_1,\varepsilon,V,-)\longmapsto_{G_1(A)}(g_1,\varepsilon,V\cdot\ass,+)
  \end{eqnarray*}
  and $f(1)=8\cdot 2^0-6=2$.
\item
  If $n>1$, then
  \begin{eqnarray*}
    (g_n,\varepsilon,V\cdot\ass,-)&\longmapsto_{G_n(A)}&(e_n,\varepsilon,V\cdot\ass\cdot\oss,-)
        \longmapsto_{G_n(A)}^{f(n-1)}(e_n,\varepsilon,V\cdot\ass\cdot\ass,+)\\
    &\longmapsto_{G_n(A)}&(h_n,\varepsilon,V\cdot\ass,-)
        \longmapsto_{G_n(A)}(j_n,\varepsilon,V,+)\longmapsto_{G_n(A)}(h_n,\varepsilon,V\cdot\oss,+)\\
    &\longmapsto_{G_n(A)}&(e_n,\varepsilon,V\cdot\oss\cdot\ass,-)
        \longmapsto_{G_n(A)}^{f(n-1)}(e_n,\varepsilon,V\cdot\oss\cdot\oss,+)\\
    &\longmapsto_{G_n(A)}&(g_n,\varepsilon,V\cdot\oss,+)\\
    (g_n,\varepsilon,V\cdot\oss,-)&\longmapsto_{G_n(A)}&(e_n,\varepsilon,V\cdot\oss\cdot\oss,-)
        \longmapsto_{G_n(A)}^{f(n-1)}(e_n,\varepsilon,V\cdot\oss\cdot\ass,+)\\
    &\longmapsto_{G_n(A)}&(h_n,\varepsilon,V\cdot\oss,-)
        \longmapsto_{G_n(A)}(j_n,\varepsilon,V,-)\longmapsto_{G_n(A)}(h_n,\varepsilon,V\cdot\ass,+)\\
    &\longmapsto_{G_n(A)}&(e_n,\varepsilon,V\cdot\ass\cdot\ass,-)
        \longmapsto_{G_n(A)}^{f(n-1)}(e_n,\varepsilon,V\cdot\ass\cdot\oss,+)\\
    &\longmapsto_{G_n(A)}&(g_n,\varepsilon,V\cdot\oss,+)
  \end{eqnarray*}
  and $f(n)=8\cdot 2^{n-1}-6=2\cdot(8\cdot 2^{n-2}-6)+6=2\cdot f(n-1)+6$.  
\end{varitemize}
In other words, proof-nets in the family $\{G_n(A)\}_{n\in\N}$ 
normalize in polynomial time but have exponentially long paths. 
The weights $W_{G_n(A)}$, as we are going to see, will be
null. This is accomplished by focusing on paths starting from
boxes, this in contrast to the execution formula~\cite{Girard89lc}, 
which takes into account conclusion-to-conclusion
paths only.\par
The second example is a 
proof-net $G$:
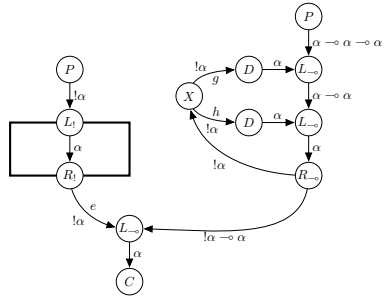
\begin{center}
    \scalebox{0.5}{\epsfbox{figure.77}}
\end{center}
Observe $G\longrightarrow^* H$
where $H$ is the following cut-free proof:
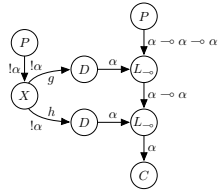
\begin{center}
    \scalebox{0.5}{\epsfbox{figure.78}}
\end{center}
The proof-net $G$ corresponds to a type
derivation for the lambda-term $(\lambda x.yxx)z$,
while $H$ corresponds to a type derivation for $yzz$.
There are finitely many paths in $C_G$, all of 
them having finite length. But the context semantics
of $G$ reflects the fact that $G$ is strongly 
normalizing in another way, too: there are finitely 
many exponential signatures $t$ such that
$(e,\varepsilon,t,+)\longmapsto_G^*(k,U,\ess,+)$, where
$k\in E_G$ and $U\in\mathscr{E}^+$. We 
can concentrate on $e$ since it is the only
box-edge on $G$. In particular:
\begin{eqnarray*}
(e,\varepsilon,\ess,+)&\longmapsto_G^*&(e,\varepsilon,\ess,+)\\
(e,\varepsilon,\rss(\ess),+)&\longmapsto_G^*&(h,\varepsilon,\ess,+)\\
(e,\varepsilon,\lss(\ess),+)&\longmapsto_G^*&(g,\varepsilon,\ess,+)
\end{eqnarray*}
Intuitively, the exponential signature $\ess$ corresponds to the initial status
of the single box in $G$, while $\lss(\ess)$ and $\rss(\ess)$ correspond to
the two copies of the same box appearing after some normalization steps.
In the following section, we will formally investigate this new way
of exploiting the context semantics as a method of studying the 
quantitative behavior of proofs.

Let us now present one last example. Consider the proof-net $J$:
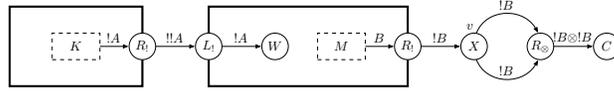
\begin{center}
    \scalebox{0.5}{\epsfbox{figure.82}}
\end{center}
The leftmost box (i.e. the box containing $K$) can interact
with the vertex $v$ and, as a consequence, can
be copied:
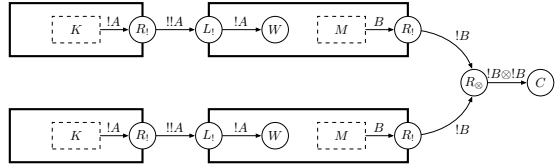
\begin{center}
    \scalebox{0.5}{\epsfbox{figure.83}}
\end{center}
However, there is not any persistent path (in the sense of ~\cite{Gonthier92lics})
between the box and $v$. The reason is simple: there is not
any \emph{path} between them. This mismatch shows why an extended
notion of path encompassing jumps between box premises and conclusions
is necessary in order to capture the quantitative behavior of proofs
(at least if \emph{every} reduction strategy is taken into account).

\section{Context Semantics and Time Complexity}\label{sect:complexity}
We are now in a position to define the weight $W_G$ of a proof-net $G$.
As already mentioned, $W_G$ takes into account the number of times each
box in $G$ is copied during normalization. Suppose $G$ contains a
sub-net matching the left-hand side of the rule $\longrightarrow_X$.
Then, there is a box-edge $e$ in $G$ such that the corresponding box
will be duplicated at least once. In the context semantics,
for every $t\in\{\ess,\lss(\ess),\rss(\ess)\}$ there are $g\in E_G$ 
and $V\in\mathscr{E}^*$ such that
\begin{equation}
(e,U,t,+)\longmapsto_G^*(g,V,\ess,b).\label{equ:copyone}
\end{equation}
As a consequence,
we would be tempted to define the ``weight'' of \emph{any} box-edge $e$ as the number
of ``maximal'' exponential signatures satisfying~(\ref{equ:copyone}). What
we need, in order to capture ``maximality'' is a notion of final contexts. 
\begin{definition}[Final Stacks, Final Contexts]
Let $G$ be a proof-net. Then:
\begin{varitemize}
\item
  First of all, we need to define what a \emph{final stack} $U\in\mathscr{S}^+$ is.
  We distinguish \emph{positive} and \emph{negative} final stacks and define
  them mutually recursively:
  \begin{varitemize}
    \item
      A positive final stack is either $\ess$ or $V\cdot \ass$ (where
      $V$ is a negative final stack) or $V\cdot s$ (where $s\in\{\oss,\fss,\xss,\sss\}$
      and $V$ is a positive final stack) or $V\cdot\ess$ (where $V$ is a positive
      final stack).
    \item
      A negative final stack is either $V\cdot \ass$ (where
      $V$ is a positive final stack) or $V\cdot s$ (where $s\in\{\oss,\fss,\xss,\sss\}$
      and $V$ is a negative final stack)  or $V\cdot t$ (where $V$ is a negative
      final stack and $t$ is an exponential signature).
  \end{varitemize}
\item
  A context $C\in C_G$ is \emph{final} iff one of the following four cases hold:
  \begin{varitemize}
    \item
      If $C=((u,v),U,V,+)$, $\alpha_G(v)=W$ and $V$ is a positive final stack;
    \item
      If $C=((u,v),U,V,+)$, $\alpha_G(v)=C$ and $V$ is a positive final stack;
    \item
      If $C=((u,v),U,\ess,+)$ and $\alpha_G(v)=D$;
    \item
      If $C=((u,v),U,V,-)$, $\alpha_G(v)=P$ and $V$ is a negative final stack;
  \end{varitemize}
\end{varitemize}
\end{definition}
Although the definition of a final stack is not trivial, the underlying
idea is very simple: if we reach a final context $C$ from $(e,U,t,+)$, then the
exponential signature $t$ must have been completely ``consumed'' along
the path. Moreover, if $C$ is final, then there are not any context $D$
such that $C\longmapsto_G D$. For example, the stack $\ess\cdot\ass\cdot\nss(\ess,\ess)$ is
negative final, while $\ess\cdot\ass\cdot\fss\cdot\ass$ is positive
final. 

Now, consider exponential signatures $t$ such that
\begin{equation}
(e,U,t,+)\longmapsto_G^*C\label{equ:copytwo}
\end{equation}
where $C$ is final. Apparently, (\ref{equ:copytwo}) could take the
place of (\ref{equ:copyone}) in defining what the weight
of any box-edge should be. However, this does not work due to rewriting rule 
$\longrightarrow_N$ which, differently from $\longrightarrow_X$, duplicates 
a \emph{box} without duplicating its \emph{content}. The binary relation 
$\sqsubseteq$ will help us to manage this mismatch. 
\begin{definition}[Copies, Canonicity, Cardinalities]
Let $G$ be a proof-net. Then:
\begin{varitemize}
\item
  A \emph{copy for $e\in B_G$ on $U\in\mathscr{E}^*$ (under $G$)}
  is a standard exponential signature $t$ such that for every $u\sqsubseteq t$ there 
  is a final context $C$ such that $(e,U,u,+)\longmapsto_G^* C$.
\item  
  A sequence $U\in\mathscr{E}^*$ is said to be \emph{canonical}
  for $e\in E_G$ iff one of the following conditions holds:
  \begin{varitemize} 
  \item
    $\theta_G(e)$ is undefined and $U$ is the
    empty sequence;
  \item
    $\theta_G(e)=v$, $V$ is canonical for $\rho_G(v)$,
    $t$ is a copy for $\rho_G(v)$ under $V$, and $U=V\cdot t$.
  \end{varitemize}
  $L_G(e)$ is the class of canonical sequences for $e$. If $v\in V_G$,
  then $L_G(v)$ is defined similarly. 
\item
  The \emph{cardinality $R_G(e,U)$ of $e\in B_G$ under $U\in\mathscr{E}^*$} is the number of different
  simplifications of copies of $e$ under $U$. $G$ has \emph{strictly positive weights} iff $R_G(e,U)\geq 1$
  whenever $U$ is canonical for $e$.
\item
  A context $(e,U,t,+)$ is said to be \emph{cyclic} for $G$
  iff $(e,U,t,+)\longmapsto_G^+(e,U,v,+)$.
\end{varitemize}
\end{definition}
Observe that $|U|=\partial(e)$ whenever $U$ is a canonical sequence for $e$.

Consider a proof-net $G$, an edge $e\in B_G$ such
that $\partial(e)=0$ and let $H$ be the box whose
conclusion is $e$. Observe that the only canonical sequence
for $e$ is $\varepsilon$. Each copy of $e$ under $\varepsilon$ corresponds to a 
potential copy of the \emph{content} of $H$.
Indeed, if $g\in B_G$, $\partial(g)=1$, and $\rho_G(g)=e$, canonical sequences 
for $g$ are precisely the copies of $e$ under $\varepsilon$.
The cardinality $R_G(e,\varepsilon)$ will be the number of
potential copies of $H$ itself, which is not necessarily 
equal to the number of copies of $e$ under $\varepsilon$: firing
a $N$-cut causes a box to be copied, without its content (see Figure~\ref{fig:exporules}).

For every proof-net $G$, $W_G$ is defined as follows:
$$
W_G=\sum_{e\in B_G}\sum_{U\in L_G(e)}(R_G(e,U)-1).
$$
The quantity $W_G$ is the \emph{weight} of the proof-net $G$. 
As we will show later, $W_G$ cannot increase
during cut-elimination. However, it is not guaranteed to
decrease at any cut-elimination step and, moreover, it is not
necessarily a bound to the size $|G|$ of $G$. As a consequence,
we need to define another quantity, called $T_G$:
$$
T_G=\sum_{v\in I_G}|L_G(v)|+\sum_{e\in B_G}P_G(e)\sum_{U\in L_G(e)}(2R_G(e,U)-1).
$$
As we will show in the following, $T_G$ is polynomially related to 
$W_G$. Since any box-edge $e\in B_G$ is charged for $P_G(e)$
in $T_G$, $T_G$ is clearly greater or equal to $|G|$. Please
notice that $W_G$ and $T_G$ can in principle be infinite.
 
We now analyze how $W_G$ and $T_G$ evolve during normalization. This will be
carried out by carefully studying to which extent paths induced by contexts semantics
are preserved during the process of cut-elimination. This task becomes easier once
the notion of canonicity is extended to contexts:
\begin{definition}[Cononical Contexts]
  A context $(e,U,V,\parity{\plength{V}{\ass}})\in C_G$ is said to be
  \emph{canonical} iff  $U$ is canonical for $e$ and
  whenever $V=W\cdot t\cdot Z$ the following
  two conditions hold:
  \begin{varnumlist}
    \item
      Either $W=\varepsilon$ and $t$ is quasi-standard or
      $W\neq\varepsilon$ and $t$ is standard.
    \item
      For every $u\sqsubseteq t$,
      it holds that $(e,U,u\cdot Z,\parity{\plength{Z}{\ass}})\longmapsto_G^* C$,
      where $C\in C_G$ is a final context.
  \end{varnumlist}
We denote with $A_G\subseteq C_G$ the set of canonical contexts.
\end{definition}
Observe, in particular, that in any canonical context
$(e,U,V,b)\in A_G$, $U$ must be canonical for $e$.
More importantly, please notice that if $t$ is a copy
for $e$ on $U$ and $U$ is canonical for $e$, then
$(e,U,t,+)$ as well as any
context $(e,U,u,+)$ (where $u\sqsubseteq t$) are canonical.
Canonicity of contexts is preserved by the relation $\longmapsto_G$:
\begin{lemma}\label{lemma:canonicalcontexts}
  If $C\in A_G$ and $C\longmapsto_G D$, then $D\in A_G$.
\end{lemma}
\begin{proof}
A straightforward case-analysis suffices. We here consider
some cases:
\begin{varitemize}
  \item
    Let $(e,U,V\cdot t,+)\longmapsto_G(g,U\cdot t,V,+)$ and suppose
    $(e,U,V\cdot t,+)$ is canonical. Clearly, $V$ must be different
    from $\varepsilon$ and, as a consequence, $t$ is standard. Moreover,
    $(e,U,u,+)\longmapsto_G^*C$ whenever $u\sqsubseteq t$. This implies
    $U\cdot t$ is a canonical sequence for $g$. Now, suppose 
    $V=W\cdot u\cdot Z$. Clearly, $u$ is quasi-standard if $W=\varepsilon$
    and standard if $W\neq\varepsilon$, because $(e,U,V\cdot t,+)$ is
    canonical. Moreover for every $v\sqsubseteq u$, either 
    $(g,U\cdot t,v\cdot Z,c)\longmapsto_G
    (e,U,v\cdot Z\cdot t,c)$ or $(e,U,v\cdot Z\cdot t,c)\longmapsto_G
    (g,U\cdot t,v\cdot Z,c)$. This implies $(g,U\cdot t,v\cdot Z,c)\longmapsto_G^* C$,
    where $C$ is final.
  \item
    Let $(e,U,V,-)\longmapsto_G(g,U,V\cdot\ass,+)$ and suppose
    $(e,U,V,-)$ is canonical. Clearly, $\theta_G(e)=\theta_G(g)$
    and, as a consequence, $U$ is canonical for $g$. Now, suppose
    $V=W\cdot t\cdot Z$. Clearly, $t$ is quasi-standard if $W=\varepsilon$
    and standard if $W\neq\varepsilon$, because $(e,U,V,-)$ is
    canonical. Moreover for every $u\sqsubseteq t$, either
    $(g,U,u\cdot Z\cdot\ass,c)\longmapsto_G
    (e,U,u\cdot Z,c\downarrow)$ or $(e,U,v\cdot Z,c)\longmapsto_G
    (g,U,u\cdot Z\cdot\ass,c\downarrow)$. This implies
    $(g,U,u\cdot Z\cdot\ass,c)\longmapsto_G^* C$,
    where $C$ is final.
  \item
    Let $(e,U,V\cdot\nss(t,u),+)\longmapsto_G(g,U,V\cdot t\cdot u,+)$ and suppose
    $(e,U,\nss(t,u),+)$ is canonical. Clearly, $\theta_G(e)=\theta_G(g)$
    and, as a consequence, $U$ is canonical for $g$. First of
    all, $u$ must be standard. Let $v\sqsubseteq u$.
    Then $\pss(v)\sqsubseteq\nss(t,u)$ and, as a consequence, 
    $(e,U,\pss(v),+)\longmapsto_G(g,U,v,+)\longmapsto_G^*C$
    where $C$ is final. $\nss(t,u)$ is standard if $V\neq\varepsilon$
    and quasi-standard if $V=\varepsilon$. As a consequence, $t$ is
    standard if $V\neq\varepsilon$ and quasi-standard if $V=\varepsilon$.
    Let $v\sqsubseteq t$. Then $\nss(v,u)\sqsubseteq\nss(t,u)$ and,
    as a consequence, $(e,U,\nss(v,u),+)\longmapsto_G(g,U,v\cdot u,+)\longmapsto_G^*C$
    where $C$ is final. Now, suppose
    $V=W\cdot v\cdot Z$. Clearly, $v$ is quasi-standard if $W=\varepsilon$
    and standard if $W\neq\varepsilon$, because $(e,U,V\cdot\nss(t,u),-)$ is
    canonical. Moreover for every $w\sqsubseteq v$, either
    $(g,U,w\cdot Z\cdot t\cdot u,c)\longmapsto_G
    (e,U,w\cdot Z\cdot\nss(t,u),c)$ or $(e,U,w\cdot Z\cdot\nss(t,u),c)\longmapsto_G
    (g,U,w\cdot Z\cdot t\cdot u,c)$. This implies $(g,U,w\cdot Z\cdot t\cdot u,c)\longmapsto_G^* C$,
    where $C$ is final. 
\end{varitemize}
This concludes the proof.
\end{proof}
As a consequence of Lemma~\ref{lemma:canonicalcontexts}, 
when analyzing how $W_G$ and $T_G$ evolve during cut-elimination
we can only assume that all involved contexts are canonical. This
will make the proofs simpler. A sequence of
canonical contexts $C_1\longmapsto_G C_2\longmapsto_G\ldots\longmapsto_G C_n$
is called a \emph{canonical path}. 

We now analyze proof-nets only containing
$W$-cuts at levels from $0$ to $n$ and $!$-cuts at level $n$.
Observe that, by definition, $G$ must be in this form
whenever $G\longrightarrowtriangle H$.
\begin{lemma}\label{lemma:weightcutfree}
Let $G$ be a proof-net, let $n\in\N$ and let $e\in B_G$ such
that $\partial(e)\leq n$. Suppose 
any cut at levels $0$ to $n-1$ in $G$ is a $W$-cut and 
any cut at level $n$ in $G$ is either a $W$-cut or a $!$-cut.
Then the only canonical context for 
$e$ is 
$$
U_e=\underbrace{\ess\cdot\ldots\cdot\ess}_{\mbox{$\partial(e)$ times}}
$$
and the only copy of $e$ on $U_e$ 
is $\ess$.
\end{lemma}
\begin{proof}
We prove the lemma by induction on $n\in\N$.
Let us first consider the case $n=0$.
We can proceed by an induction on the structure of a proof $\pi$ such
that $G=G_\pi$, here. The only interesting inductive case is the one for the rule
corresponding to $U$. By hypothesis, it must be either a $W$-cut or
a $!$-cut. The case $n>0$ can be treated in the same way. This concludes the proof.
\end{proof}
As a consequence, any proof-net $G$ satisfying the conditions of Lemma~\ref{lemma:weightcutfree}
has strictly positive weights, although $W_G=0$. We can go even further, proving that
$A_G$ does not contain any cycle whenever $G$ only contains $W$-cuts:
\begin{lemma}\label{lemma:abscyclecutfree}
Let $G$ be a proof-net containing $W$-cuts only. Then $A_G$ does not contain any
cycle.
\end{lemma}
\begin{proof}
We can prove the following, stronger statement by a straightforward induction on $G$:
if $(e,U,V,b)\mapsto_G^+(e,W,Z,c)$ then $b\neq c$.
\end{proof}
If $G\Longrightarrow_S H$, the property of having strictly positive weights and
not containing canonical cycles propagates from $H$ to $G$. Moreover, it is possible
to precisely evaluated the difference between $W_G$ and $W_H$, depending on $S$.
Independently on $S$, $T_G$ is going to be strictly higher that $T_H$. Formally:
\begin{lemma}\label{lemma:monotonicityaux}
Suppose that $G\Longrightarrow_S H$, $H$ has strictly positive weights and 
$A_H$ does not contain any cicle. Then:
\begin{varitemize}
  \item
  $G$ has strictly positive weights;
  \item
  $A_G$ does not contain any cycle;
  \item 
  $T_G>T_H$;
  \item
  If $S\in\{\linear,\otimes,\forall,D,W\}$, then $W_G=W_H$;
  \item
  If $S=!$, then $W_G=W_H+\sum_{U\in L_G(e)}R_G(e,U)$, where $e$ is the
  box edge involved in the cut-elimination step;
  \item
  If $S\in\{X,N\}$, then $W_G=W_H+|L_G(e)|$, where $e$ is the box edge involved
  in the cut-elimination step.
\end{varitemize}
\end{lemma}
\begin{proof}
We can distinguish some cases:
\begin{varitemize}
  \item
    Let now $G\Longrightarrow_\linear H$. Then we are in the following situation:
    \begin{center}
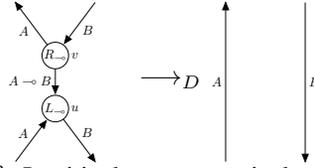
      \begin{minipage}[c]{61pt}
        \centering\scalebox{0.5}{\epsfbox{figure.67}}
      \end{minipage}
      \begin{minipage}[c]{20pt}
        \centering $\longrightarrow_D$
      \end{minipage}
      \begin{minipage}[c]{51pt}
        \centering\scalebox{0.5}{\epsfbox{figure.68}}
      \end{minipage}
    \end{center}
    Observe that $B_G=B_H$ and $I_G=I_H\cup\{u,v\}$. Intuitively, any canonical path in
    $G$ can be mimicked by a canonical path in $H$ and viceversa. We can make this
    claim more precise: for every $e\in B_G=B_H$, $L_G(e)=L_H(e)$ and, moreover,
    for every $U\in L_G(e)=L_H(e)$ and for every
    $t\in\mathscr{E}$, $t$ is a copy for $e$ on $U$ under $G$ iff
    $t$ is a copy for $e$ on $U$ under $H$. We can proceed by induction on 
    $\partial(e)$:
    \begin{varitemize}
    \item
      If $\partial(e)$=0, then by definition $L_G(e)=L_H(e)=\{\varepsilon\}$.
      Moreover, for every exponential signature $u$ and every final context
      $C$ (for $G$ or for $H$), we have $(e,\varepsilon,u,+)\longmapsto_G^* C$ iff 
      $(e,\varepsilon,u,+)\longmapsto_H^* C$. This implies the thesis.
    \item
      If $\partial(e)>0$, then $\rho_G(e)=\rho_H(e)$ and
      $\partial(\rho_G(e))=\partial(\rho_H(e))<\partial(e)$. By the inductive
      hypothesis, $L_G(\rho_G(e))=L_H(\rho_H(e))$. This implies
      $L_G(e)=L_H(e)$, because elements of $L_G(e)$ are defined 
      by extending $U\in L_G(\rho_G(e))$ with a copy 
      for $\rho_G(e)$ on $U$ (and the same definition applies to $H$).
      Moreover, for every exponential signature $u$, every $U\in L_G(e)=L_H(e)$ 
      and every final context $C$ (for $G$ or for $H$), 
      we have $(e,U,u,+)\longmapsto_G^* C$ iff 
      $(e,U,u,+)\longmapsto_H^* C$. This implies the thesis.
    \end{varitemize}
    As a consequence, $R_G(e,U)=R_H(e,U)$.
    whenever $e$ is a box edge and whenever $U$ is canonical for $e$. This implies
    $W_G=W_H$. Moreover $L_G(w)=L_H(w)$ whenever $w\in V_G$. But
    since $|L_G(v)|,|L_G(u)|\geq 1$, we have $T_G>T_H$. If $H$
    is strictly positive then $H$ is strictly positive, too.
    Finally, if $A_G$ contains a cycle for $G$, then this same cycle is 
    a cycle in $A_H$ for $H$.
  \item
    Let $G\Longrightarrow_\forall H$. Then we are in the following situation:
    \begin{center}
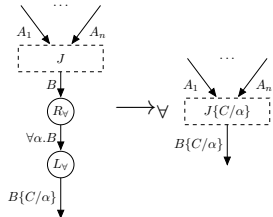
      \begin{minipage}[c]{38pt}
        \centering\scalebox{0.5}{\epsfbox{figure.71}}
      \end{minipage}
      \begin{minipage}[c]{20pt}
        \centering $\longrightarrow_\forall$
      \end{minipage}
      \begin{minipage}[c]{38pt}
        \centering\scalebox{0.5}{\epsfbox{figure.72}}
      \end{minipage}
    \end{center}
    The argument used in the previous
    case applies here, too. Notice that $J\{C/\alpha\}$ is structurally identical to
    $J$ (they only differs in the labelling functions $\beta_J$ and $\beta_{J\{C/\alpha\}}$).
    We can conclude that $W_G=W_H$, $T_G>T_H$, $G$ is strictly positive whenever
    $H$ is and if $A_G$ contains a cycle, then $A_H$ contains a cycle, too.
  \item
    Let $G\Longrightarrow_! H$. Then we are in the following situation:
    \begin{center}
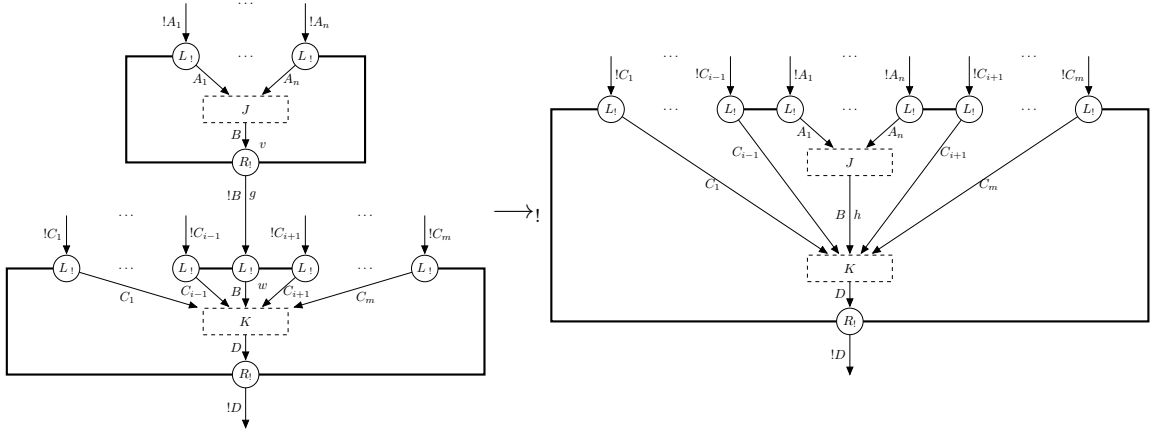
      \begin{minipage}[c]{181pt}
        \centering\scalebox{0.5}{\epsfbox{figure.73}}
      \end{minipage}
      \begin{minipage}[c]{20pt}
        \centering $\longrightarrow_!$
      \end{minipage}
      \begin{minipage}[c]{226pt}
        \centering\scalebox{0.5}{\epsfbox{figure.74}}
      \end{minipage}
    \end{center}
    Observe that $I_G=I_H$ and $B_G=B_H\cup\{g\}$.
    By the same induction methodology we used in the
    first case we can prove the following: for every 
    $e\in B_H$, $L_G(e)=L_H(e)$ and, moreover,
    for every $e\in B_H$, for every $U\in L_H(e)=L_G(e)$ and for every
    $t\in\mathscr{E}$, $t$ is a copy for $e$ on $U$ under $G$ iff
    $t$ is a copy for $e$ on $U$ under $H$. Observe that here proving
    the preservation of paths become a bit more delicate and,
    in particular, Lemma~\ref{lemma:canonicalcontexts} is crucial.
    For example, suppose we want to mimick a canonical path in $H$ going 
    from $K$ to $J$ through $h$ by a path in $G$ going through $g$. 
    This can be done only if any context $C=(r,U,V,c)$ is 
    such that $U\neq\varepsilon$. But since we know that 
    $C\in A_H$, we can conclude that, indeed,
    $U$ is canonical for $h$, and is nonempty.
    This implies that $R_G(h,U)$ is always equal to $R_H(h,U)$ except
    when $h=e$. As a consequence, $T_G>T_H$ and
    $W_G=W_H+\sum_{U\in L_G(g)}R(g,U)$. Notice that 
    $\sum_{U\in L_G(g)}R(g,U)=1$ whenever $G\longrightarrowtriangle H$.
    If $A_G$ contains a cycle, then $A_H$ contains a cycle, too.
  \item
    Let $G\Longrightarrow_D H$. Then we are in the following situation:
    \begin{center}
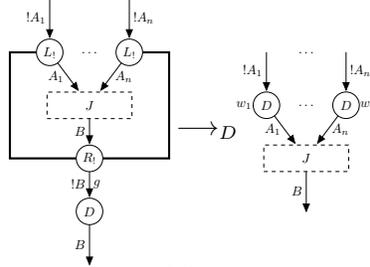
      \begin{minipage}[c]{61pt}
        \centering\scalebox{0.5}{\epsfbox{figure.61}}
      \end{minipage}
      \begin{minipage}[c]{20pt}
        \centering $\longrightarrow_D$
      \end{minipage}
      \begin{minipage}[c]{51pt}
        \centering\scalebox{0.5}{\epsfbox{figure.62}}
      \end{minipage}
    \end{center}
    Observe that $I_G\supseteq I_H-\{w_1,\ldots,w_n\}$ and $B_G=B_H\cup\{g\}$. 
    Furthermore, notice that $R_G(g,U)=1$ for every $U$, since
    the only copy of $g$ on any $U$ is $\ess$. We can prove the
    following for every $e\in B_H$:
    \begin{varitemize}
    \item
      If $e\in B_J$, then
      $$
      L_G(e)=\{U\cdot \ess\cdot V\;|\;U\cdot V\in L_H(e)\mbox{ and }|U|=\partial(e)\}
      $$
      and, moreover, for every $U\cdot \ess\cdot V\in L_G(e)$ (where
      $|U|=\partial(e)$) and for every $t\in\mathscr{E}$, $t$ is a copy for $e$ 
      on $U\cdot \ess\cdot V$ under $G$ iff
      $t$ is a copy for $e$ on $U\cdot V$ under $H$.
    \item
      If $e\notin B_J$, then $L_G(e)=L_H(e)$ and, moreover,
      for every $U\in L_G(e)$ and for every $t\in\mathscr{E}$,
      $t$ is a copy for $e$ 
      on $U$ under $G$ iff
      $t$ is a copy for $e$ on $U$ under $H$.
    \end{varitemize}
    As usual, we can proceed by induction. 
    As a consequence, $W_G=W_H$. 
    Notice that $n=P_G(g)-1$. This implies
    $T_G>T_G$, since $|L_G(w_i)|=|L_G(g)|$. If $A_G$ contains a cycle, a cycle can
    be found in $A_H$ as well.
  \item
    Let $G\Longrightarrow_W H$. Then $G$ and $H$ only contain $W$-cuts.
    By lemma~\ref{lemma:weightcutfree}, $G$ and $H$ satisfy claim~\ref{claim:posweights}. 
    By lemma~\ref{lemma:abscyclecutfree}, $G$ and $H$ satisfy claim~\ref{claim:abscycles}.
    Moreover, $W_G=W_H=0$. We are in the following situation:
    \begin{center}
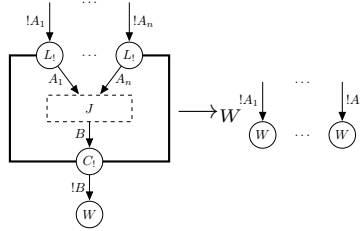
      \begin{minipage}[c]{61pt}
        \centering\scalebox{0.5}{\epsfbox{figure.75}}
      \end{minipage}
      \begin{minipage}[c]{20pt}
        \centering $\longrightarrow_W$
      \end{minipage}
      \begin{minipage}[c]{51pt}
        \centering\scalebox{0.5}{\epsfbox{figure.76}}
      \end{minipage}
    \end{center}
    But
    \begin{eqnarray*}
      T_G&=&\sum_{v\in I_G}|L_G(v)|+\sum_{e\in B_G}P_G(e)\sum_{U\in L_G(e)}(2R_G(e,U)-1)\\
      &=&\sum_{v\in I_G}1+\sum_{e\in B_G}P_G(e)\cdot 1\\
      &=&|I_G|+\sum_{e\in B_G}P_G(e)\\
      &\geq&|I_H|-n+\sum_{e\in B_H}P_H(e)+P_G(g)\\
      &=&|I_H|-n+\sum_{e\in B_H}P_H(e)+n+1\\
      &>&|I_H|+\sum_{e\in B_H}P_H(e)\\
      &=&\sum_{v\in I_H}|L_H(v)|+\sum_{e\in B_H}P_H(e)\sum_{U\in L_H(e)}(2R_H(e,U)-1)\\
      &=&T_H
    \end{eqnarray*}
  \item
    Suppose $G\Longrightarrow_X H$. Then we are in the following situation:
    \begin{center}
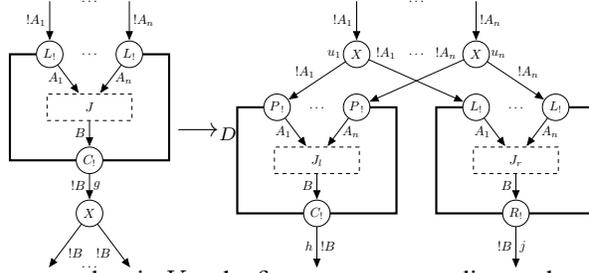
      \begin{minipage}[c]{61pt}
        \centering\scalebox{0.5}{\epsfbox{figure.63}}
      \end{minipage}
      \begin{minipage}[c]{20pt}
        \centering $\longrightarrow_D$
      \end{minipage}
      \begin{minipage}[c]{51pt}
        \centering\scalebox{0.5}{\epsfbox{figure.64}}
      \end{minipage}
    \end{center}
    For every edge $e\in V_J$, there are two edges
    $e_l$ and $e_j$ in $V_H$, the first one corresponding
    to the copy of $e$ in $J_l$ and the second one corresponding
    to the copy of $e$ in $J_r$. We can prove the following for every $e\in B_G$:
    \begin{varitemize}
    \item
      If $e\in B_J$, then
      $$
      L_G(e)=\{U\cdot \lss(t)\cdot V\;|\;U\cdot t\cdot V\in L_H(e_l)\mbox{ and }|U|=\partial(e)\}
      \cup\{U\cdot \rss(t)\cdot V\;|\;U\cdot t\cdot V\in L_H(e_r)\mbox{ and }|U|=\partial(e)\}.
      $$
      Moreover, for every $U\cdot \lss(u)\cdot V\in L_G(e)$ (where
      $|U|=\partial(e)$) and for every $t\in\mathscr{E}$, $t$ is a copy for $e$ 
      on $U\cdot \lss(u)\cdot V$ under $G$ iff
      $t$ is a copy for $e_l$ on $U\cdot u\cdot V$ under $H$.
      Furthermore, for every $U\cdot \rss(u)\cdot V\in L_G(e)$ (where
      $|U|=\partial(e)$) and for every $t\in\mathscr{E}$, $t$ is a copy for $e$ 
      on $U\cdot \rss(u)\cdot V$ under $G$ iff
      $t$ is a copy for $e_r$ on $U\cdot u\cdot V$ under $H$.
    \item
      If $e\notin B_J$ and $e\neq g$, then $L_G(e)=L_H(e)$ and, moreover,
      for every $U\in L_G(e)$ and for every $t\in\mathscr{E}$,
      $t$ is a copy for $e$ 
      on $U$ under $G$ iff
      $t$ is a copy for $e$ on $U$ under $H$.
    \item
      $L_G(g)=L_H(h)=L_H(j)$ and 
      for every $U\in L_G(g)$ and for every $t\in\mathscr{E}$,
      $t$ is a copy for $e$ 
      on $U$ under $G$ iff
      $t=\lss(u)$ and $u$ is a copy for $r$ on $U$ under $H$ or
      $t=\rss(u)$ and $u$ is a copy for $q$ on $U$ under $H$.
    \end{varitemize}
    As usual, we can proceed by induction on $\partial(e)$
    and Lemma~\ref{lemma:canonicalcontexts} is crucial. 
    It follows that $R_G(g,U)=R_H(h,U)+R_H(j,U)$ and $W_G=W_H+|L_G(g)|$. 
    Moreover, notice that for every vertex $w\in I_J$, 
    there are two vertices $z\in I_{J_l}$
    and $s\in I_{J_r}$ such that $|L_G(w)|=|L_H(z)|+|L_H(s)|$. 
    Since $n=P_G(g)-1$, we can conclude
    $T_G>T_H$. If $A_G$ contains a cycle, a cycle can be found in $A_H$, too.
  \item
    Suppose $G\longrightarrow_N H$. Then we are in the following situation:
    \begin{center} 
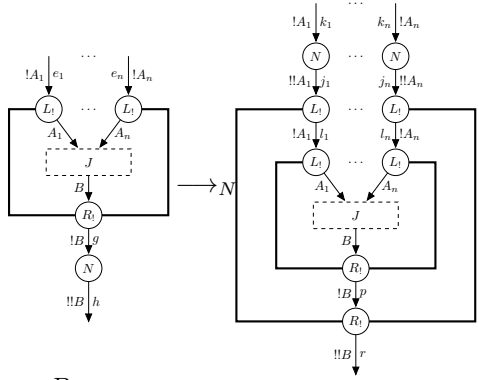
      \begin{minipage}[c]{61pt}
        \centering\scalebox{0.5}{\epsfbox{figure.65}}
      \end{minipage}
      \begin{minipage}[c]{20pt}
        \centering $\longrightarrow_N$
      \end{minipage}
      \begin{minipage}[c]{91pt}
        \centering\scalebox{0.5}{\epsfbox{figure.66}}
      \end{minipage}
    \end{center}
 We can prove the following for every $e\in B_G$:
    \begin{varitemize}
    \item
      If $e\in B_J$, then
      $$
      L_G(e)=\{U\cdot \nss(u,v)\cdot V\;|\;U\cdot v\cdot u\cdot V\in L_H(e)\mbox{ and }|U|=\partial(e)\}.
      $$
      Moreover, for every $U\cdot \nss(u,v)\cdot V\in L_G(e)$ (where
      $|U|=\partial(e)$) and for every $t\in\mathscr{E}$, $t$ is a copy for $e$ 
      on $U\cdot \nss(u,v)\cdot V$ under $G$ iff
      $t$ is a copy for $e_l$ on $U\cdot v\cdot u\cdot V$ under $H$.
    \item
      If $e\notin B_J$ and $e\neq g$, then $L_G(e)=L_H(e)$ and, moreover,
      for every $U\in L_G(e)$ and for every $t\in\mathscr{E}$,
      $t$ is a copy for $e$ 
      on $U$ under $G$ iff
      $t$ is a copy for $e$ on $U$ under $H$.
    \item
      $L_G(g)=L_H(j)$ and 
      for every $U\in L_G(g)$ and for every $t\in\mathscr{E}$,
      $t$ is a copy for $e$ 
      on $U$ under $G$ iff
      $t=\nss(u,v)$, $v$ is a copy for $j$ on $U$ under $H$ 
      and $u$ is a copy for $h$ on $v\cdot U$ under $H$.
    \end{varitemize}
    As usual, we can proceed by induction on $\partial(e)$
    and Lemma~\ref{lemma:canonicalcontexts} is crucial. 
    But notice that a simplification
    of $t$ is either $\nss(w,v)$, where $w$ is a simplification of $u$ or
    $\pss(z)$ where $z$ is a simplification of $v$. It follows that 
    $$
    R_G(g,U)=R_H(j,U)+\sum_{t\cdot U\in L_H(j)} R_H(h,t\cdot U),
    $$  
    and, as a consequence, $W_G=W_H+|L_G(g)|$.
    Moreover, notice that for every vertex $w$ in $J$, 
    it holds that $|L_G(w)|=|L_H(w)|$. Since $n=P_G(g)-1$, we 
    can conclude $T_H\leq T_G+|L_G(g)|(P_G(g)-1)$.
\end{varitemize}
This concludes the proof.
\end{proof}
Lemma~\ref{lemma:monotonicityaux} gives us enough information to
establish strong correspondences between $T_G$, $W_G$ and the number of
steps necessary to rewrite $G$ to normal form:
\begin{proposition}[Positive Weights, Absense of Cycles and Monotonicity]\label{prop:normlinear}
Let $G$ be a proof-net. Then
\begin{varnumlist}
\item\label{claim:posweights}
  $G$ has strictly positive weights;
\item\label{claim:abscycles}
  $A_G$ does not contain any cycle;
\item\label{claim:decreasing}
  $W_G\geq W_H$ and $T_G>T_H$ whenever $G\Longrightarrow H$;
\item\label{claim:slowdecreasing}
  $W_G\leq W_H+1$ whenever $G\longrightarrowtriangle H$,
\end{varnumlist}
\end{proposition}
\begin{proof}
We prove claims~\ref{claim:posweights} to~\ref{claim:decreasing}
by induction on $\stepsto{G}{\Longrightarrow}$ and claim~\ref{claim:slowdecreasing}
by induction on $\stepsto{G}{\longrightarrowtriangle}$. First of all, consider
a proof-net $G$ such that $\stepsto{G}{\Longrightarrow}=0$. 
Clearly, $G$ must be cut-free. By lemma~\ref{lemma:weightcutfree},
$G$ satisfies claim~\ref{claim:posweights}. By lemma~\ref{lemma:abscyclecutfree},
$G$ satisfies claim~\ref{claim:abscycles}. Moreover, there cannot be 
any $H$ such that $G\Longrightarrow H$.
Now, suppose $[G]_{\Longrightarrow}\geq 1$ and suppose
$G\Longrightarrow H$. Clearly $[H]_{\Longrightarrow}<[G]_{\Longrightarrow}$
and, as a consequence, we can assume $H$ satisfies conditions~\ref{claim:posweights}
and~\ref{claim:abscycles}. We can prove $G$ satisfies conditions~\ref{claim:posweights} 
to~\ref{claim:decreasing} by lemma~\ref{lemma:monotonicityaux}.

Now, consider a proof-net $G$ such that $\stepsto{G}{\longrightarrowtriangle}=0$. 
Clearly, $G$ must be cut-free. As a consequence, there cannot be 
any $H$ such that $G\longrightarrowtriangle H$ and condition~\ref{claim:slowdecreasing}
is satisfied. Now, suppose $\stepsto{G}{\longrightarrowtriangle}\geq 1$ and let
$G\longrightarrowtriangle_S H$. Clearly, we can assume
$G$ satisfies conditions~\ref{claim:posweights} 
to~\ref{claim:slowdecreasing}. From lemma~\ref{lemma:monotonicityaux}, we know
that $W_G=W_H$ if $S\in\{\linear,\otimes,\forall,D,W\}$. Suppose
$S\in\{!,X,N\}$ and let $e\in B_G$ be the cut-edge involved in the
cut-elimination step. From lemma~\ref{lemma:weightcutfree}, 
we know that $L_G(e)=\{U_e\}$ and $R_G(e,U_e)=1$. By lemma~\ref{lemma:monotonicityaux},
this implies the thesis.
\end{proof}
The following is a technical lemm that will be essential in proving $T_G$ to
b polynomially related to $W_G$:
\begin{lemma}\label{lemma:bound}
Let $G$ be a proof-net and let $e\in B_G$. Then,
$\sum_{U\in L_G(e)}R_G(e,U)\leq W_G+1$.
\end{lemma}
\begin{proof}
Let $D_G(e)\subseteq B_G$ be defined as follows:
$$
D_G(e)=\left\{
\begin{array}{ll}
\{e\}\cup D_G(\sigma_G(e)) & \mbox{if $\sigma_G(e)$ is defined}\\
\{e\}& \mbox{otherwise}
\end{array}\right.
$$ 
We will prove the following statement
$$
\sum_{U\in L_G(e)}R_G(e,U)\leq\left(\sum_{g\in D_G(e)}\sum_{U\in L_G(g)}(R_G(g,U)-1)\right)+1.
$$
We go by induction on $\partial(e)$. If $\partial(e)=0$, then $L_G(e)=\{\varepsilon\}$
and $D_G(e)=\{e\}$. Then
$$
\sum_{U\in L_G(e)}R_G(e,U)=R_G(e,\varepsilon)=R_G(e,\varepsilon)-1+1=
\left(\sum_{g\in D_G(e)}\sum_{U\in L_G(g)}(R_G(g,U)-1)\right)+1.
$$
 If $\partial(e)>0$, then $\sigma_G(e)$ is defined and, moreover,
\begin{eqnarray*}
\sum_{U\in L_G(e)}R_G(e,U)&=&\left(\sum_{U\in L_G(e)}(R_G(e,U)-1)\right)+|L_G(e)|\\
  &\leq&\left(\sum_{U\in L_G(e)}(R_G(e,U)-1)\right)+\sum_{U\in L_G(\sigma_G(e))}R_G(e,U)\\
  &\leq&\left(\sum_{U\in L_G(e)}(R_G(e,U)-1)\right)+
   \left(\sum_{g\in D_G(\sigma_G(e))}\sum_{U\in L_G(g)}(R_G(g,U)-1)\right)+1\\
  &=&\sum_{g\in D_G(e)}\sum_{U\in L_G(g)}(R_G(g,U)-1)+1.
\end{eqnarray*}
Now observe that for every $e\in B_G$, $D_G(e)\subseteq B_G$ and,
as a consequence, 
$$
\left(\sum_{g\in D_G(e)}\sum_{U\in L_G(g)}(R_G(g,U)-1)\right)+1\leq W_G+1.
$$
This concludes the proof.
\end{proof}
As a consequence of Proposition~\ref{prop:normlinear}, $T_G$ bounds the number of cut-elimination
steps necessary to rewrite $G$ to its normal form. As it can be easily shown, $T_G$ is also an upper
bound on $|G|$. The following result can then be obtained by proving appropriate inequalities
between $W_G$, $|G|$ and $T_G$:
\begin{theorem}\label{theo:first}
There is a polynomial $p:\N^2\rightarrow\N$ such that
for every proof-net $G$, $\stepsto{G}{\longrightarrow},\dlength{G}{\longrightarrow}\leq p(W_G,|G|)$.
\end{theorem}
\begin{proof}
By Proposition~\ref{prop:normlinear}, we can conclude
that  $T_G>T_H$ whenever $G\Longrightarrow H$. Moreover, 
by lemma~\ref{lemma:bound},
\begin{eqnarray*}
T_G&=&\sum_{e\in B_G}P_G(e)\sum_{U\in L_G(e)}(2R_G(e,U)-1)+\sum_{v\in I_G}|L_G(v)|\\
&\leq&\sum_{e\in B_G}2|G|\sum_{U\in L_G(e)}R_G(e,U)
  +\sum_{v\in I_G}(W_G+1)\\
&\leq& \sum_{e\in B_G}2|G|(W_G+1)+|G|(W_G+1)\\
&\leq& 2|G|^2(W_G+1)+|G|(W_G+1)\\
&=& (2|G|^2+|G|)(W_G+1)
\end{eqnarray*}
Finally:
$$
T_G\geq\sum_{e\in B_G}P_G(e)+|I_G|=|V_G|=|G|
$$
Since $T_G\geq 0$ for every $G$, it is clear
that $\stepsto{G}{\Longrightarrow},\dlength{G}{\Longrightarrow}\leq p(W_G,|G|)$,
where $p(x,y)=(2y^2+y)(x+1)$. This concludes the proof,
since, by lemma~\ref{lemma:standardization}, $\stepsto{G}{\longrightarrow}=\stepsto{G}{\Longrightarrow}$
and $\dlength{G}{\longrightarrow}=\dlength{G}{\Longrightarrow}$.
\end{proof}
The weight $W_G$ can only decrease during cut-elimination. Moreover, it decreases by at
most one at any normalization step when performing the level-by-level strategy. As a 
consequence, the following theorem holds:
\begin{theorem}\label{theo:second}
Let $G$ be a proof-net. There is $H$ 
with $G\longrightarrowtriangle^{W_G} H$.
\end{theorem}
\begin{proof}
By Proposition~\ref{prop:normlinear}, $W_G$ decreases by at most one at any
normalization step when performing the ``level-by-level'' strategy $\longrightarrowtriangle$.
Observe that $W_G=0$ whenever $G$ is cut-free. This concludes the proof.
\end{proof}
Theorems~\ref{theo:first} and~\ref{theo:second} highlights the existence
of strong relations between context semantics and computational complexity. 
The two results can together be seen as a strengthening of the well-known correspondence
between strongly normalizing nets and finiteness of regular paths 
(see~\cite{danos95advances}). This has very interesting consequences:
for example, a family $\mathscr{G}$ of proof-nets can be normalized
in polynomial (respectively, elementary) time iff there is a polynomial
(respectively, an elementary function) $p$ such that $W_{G}\leq p(|G|)$ 
for every $G\in\mathscr{G}$. This will greatly help in the following 
section, where we sketch new proofs of soundness for various subsystems 
of linear logic.


Now, suppose $t$ is a copy of $e\in B_G$ under $U\in\mathscr{E}^*$.
By definition, there is a finite (possibly empty) sequence
$C_1,\ldots,C_n$ such that
$$
(e,U,t,+)\longmapsto_G C_1\longmapsto_G C_2\longmapsto_G\ldots\longmapsto_G C_n
$$
and $C_n$ is final. But what else can be said about this sequence?
Let $C_i=(g_i,V_i,W_i,b_i)$ for every $i$. By induction on $i$, the leftmost
component of $W_i$ must be an exponential signature, i.e.\ $W_i=u_i\cdot Z_i$
for every $i$. Moreover, every $u_i$ must be a subtree of $t$ (another
easy induction on $i$). This observation can in fact be slightly generalized into
the following result:
\begin{proposition}[Subtree Property]
Suppose $t$ is a standard exponential signature. For every subtree $u$ of
$t$, there is $v\sqsubseteq t$ such that, whenever $G$ is a proof-net,
$U\in\mathscr{E}^*$ is canonical for $e\in B_G$ and $t$ is a copy of
$e$ on $U$, there are $g\in E_G$ and $V\in\mathscr{E}^*$ with
$(e,U,v,+)\longmapsto_G^*(g,V,u,+)$.
\end{proposition}
\begin{proof}
We prove the following, stronger statement: for every exponential signature
$t$ and for every subtree $u$ of $t$, there is $v\sqsubseteq t$ such
that whenever $(e,U,t,+)\in A_G$, there are $g\in E_G$ and $W\in\mathscr{E}^*$
with $(e,U,v,+)\longmapsto_G^*(g,V,u,+)$. We proceed by induction on $t$:
\begin{varitemize}
  \item
    If $t=\ess$, then $g=e$ and $V=U$.
  \item
    If $t=\rss(w)$, then $u=t$ or $u$ is a subtree of $w$. In the first
    case $g=e$ and $V=U$. In the second case, apply the induction hypothesis
    to $w$ and $u$ obtaining a term $z\sqsubseteq w$. Since 
    $(e,U,\rss(w),+)\longmapsto_G^* C$ and  $(e,U,\rss(z),+)\longmapsto_G^* D$
    where $C,D$ are final, we can conclude that 
    \begin{eqnarray*}
      (e,U,\rss(w),+)&\longmapsto_G^*& (h,W,w,+)\\
      (e,U,\rss(z),+)&\longmapsto_G^*& (h,W,z,+)
    \end{eqnarray*}
    \begin{sloppypar}
    for some $h,W$. By induction hypothesis, $(h,W,z,+)\longmapsto_G^*(g,V,u,+)$ 
    for some $g,V$ and, as a consequence $(e,U,\rss(z),+)\longmapsto_G^*(g,V,u,+)$.
    \end{sloppypar}
  \item
    If $t=\lss(w)$ or $t=\pss(w)$ then we can proceed as in the preceeding
    case.
  \item
    If $t=\nss(w,z)$, then $u=t$ or $t$ is a subtree of $z$ or $t$
    is a subtree of $w$. In the first case, $g=e$ and $V=U$ as usual.
    In the second case, apply the induction hypothesis to
    $z$ and $u$ obtaining a term $x\sqsubseteq z$. Notice that
    $\pss(x)\sqsubseteq\nss(w,z)$ and $\pss(z)\sqsubseteq\nss(w,z)$.
    Since $(e,U,\pss(x),+)\longmapsto_G^* C$ and  $(e,U,\pss(z),+)\longmapsto_G^* D$
    where $C,D$ are final, we can conclude that 
    \begin{eqnarray*}
      (e,U,\pss(z),+)&\longmapsto_G^*& (h,W,z,+)\\
      (e,U,\pss(x),+)&\longmapsto_G^*& (h,W,x,+)
    \end{eqnarray*}
    \begin{sloppypar}
    for some $h,W$. By induction hypothesis, $(h,W,x,+)\longmapsto_G^*(g,V,u,+)$
    for some $g,V$ and, as a consequence $(e,U,\pss(x),+)\longmapsto_G^*(g,V,u,+)$.
    \end{sloppypar}
    In the third case, we can assume $u\neq\ess$ and apply the 
    induction hypothesis to $w$ and $u$ obtaining a term $y\sqsubseteq z$.
    Notice that $\nss(y,z)\sqsubseteq\nss(w,z)$. Since
    $(e,U,\nss(y,z),+)\longmapsto_G^* C$ and  $(e,U,\nss(w,z),+)\longmapsto_G^* D$
    where $C,D$ are final and $y,w\neq\ess$, we can conclude that
   \begin{eqnarray*}
      (e,U,\nss(y,z),+)&\longmapsto_G^*& (h,W,y,+)\\
      (e,U,\nss(w,z),+)&\longmapsto_G^*& (h,W,w,+)
    \end{eqnarray*}
    \begin{sloppypar}
    for some $h,W$. By induction hypothesis, $(h,W,y,+)\longmapsto_G^*(g,V,u,+)$
    for some $g,V$ and, as a consequence $(e,U,\nss(y,z),+)\longmapsto_G^*(g,V,u,+)$.
  \end{sloppypar}
\end{varitemize}
This concludes the proof.
\end{proof}
The subtree property is extremely useful when proving bounds on $R_G(e,U)$ and
$W_G$ in subsystems of \MELL. The intuitive idea behind the subtree property
is the following: whenever $t$ is a copy of $e$ under $U$ and $U$ is canonical 
for $e$, the exponenial signature $t$ must be completely 
``consumed'' along the canonical path leading from $(e,U,t,+)$ to a 
final context $C$.

\section{Subsystems}\label{sect:subsystems}
In this section, we will give some arguments about the 
usefulness of context semantics by analyzing three subsystems
of \MELL\ from a complexity viewpoint.\par
\subsection{Elementary Linear Logic}
Elementary linear logic (\ELL,~\cite{Girard98ic}) is
just \MELL\ with a weaker modality: rules $D_!$ and $N_!$
are not part of the underlying sequent calculus.
This restriction enforces the following property at
the semantic level:
\begin{lemma}[Stratification]
Let $G$ be a \emph{\ELL} proof-net. 
If $(e,U,V,b)\longmapsto^*_G(g,W,Z,c)$,
then $\tlength{U}+\tlength{V}=\tlength{W}+\tlength{Z}$.
\end{lemma}
\begin{proof}
Suppose $(e,U,V,b)\longmapsto_G^n(g,W,Z,c)$, where $n\geq 0$. By
induction on $n$, we can prove that $\tlength{U}+\tlength{V}=
\tlength{W}+\tlength{Z}$. Notice that the only rewriting
rules that can break the above equality in \MELL\ are precisely those
induced by $D$ and $N$.
\end{proof}
By exploiting stratification together with the subtree property, we
can easily prove the following result:
\begin{proposition}[\ELL\ Soundness]\label{prop:ELLsound}
For every $n\in\N$ there is an elementary
function $p_n:\N\rightarrow\N$ such that
$W_G\leq p_{\partial(G)}(\length{G})$ for every \emph{\ELL} 
proof-net $G$.
\end{proposition}
\begin{proof}
For every $n\in\N$, define two elementary functions $r_n,q_n:\N\rightarrow\N$
as follows:
\begin{eqnarray*}
\forall x.r_0(x)&=&1;\\
\forall n.\forall x.q_n(x)&=&2^{x\cdot r_n(x)+1};\\
\forall n.\forall x.r_{n+1}(x)&=&r_n(x) q_n(x).
\end{eqnarray*}
We can now prove that for every $e\in B_G$ and whenever $U$ is canonical for $e$
the following inequalities hold:
\begin{eqnarray*}
|L_G(e)|&\leq& r_{\partial(e)}(|G|)\\
R_G(e,U)&\leq& q_{\partial(e)}(|G|)
\end{eqnarray*}
We can proceed by induction on $\partial(e)$. If $\partial(e)=0$, then the only canonical
sequence for $e$ is $\varepsilon$ and the first inequality is satisfied. Moreover,
any copy of $e$ under $\varepsilon$ is an exponential signature containing at
most $|G|$ instances of $\rss$ and $\lss$ constructors: by way of contraddiction,
suppose $t$ is a copy of $e$ under $\varepsilon$ containing $m>|G|$ constructors.
Then, by the subtree property, there are $m$ distinct subterms $u_1,\ldots,u_m$
of $t$ and $g_1,\ldots,g_m\in E_G$ such that $(e,\varepsilon,t,+)\longmapsto_G^*(g_i,\varepsilon,u_i,+)$.
for every $i$. Clearly, $g_i=g_j$ for some $i\neq j$ (since $m>|G|$ and
the $g_i$ can always be chosen as to be the only edge incident to a vertex
labelled with $X$, $C$, $W$ or the only edge leaving from a vertex labelled
with $P$), but this contraddicts acyclicity.
As a consequence, the second inequality is satisfied, because there are
at most $2^{|G|+1}$ exponential signatures with length at most $|G|$. If $\partial(e)>0$,
we can observe that canonical sequences for $\partial(e)$ are in the form $V\cdot t$,
where $V$ is canonical for $\sigma_G(e)$ and $t$ is a copy for $\sigma_G(e)$ under $V$.
By the induction hypothesis we can conclude that:
\begin{eqnarray*}
|L_G(e)|&\leq&\sum_{U\in L_G(\sigma_G(e))}R_G(\sigma_G(e),U)\leq\sum_{U\in L_G(\sigma_G(e))}q_{\partial(e)-1}(|G|)\\
   &\leq& r_{\partial(e)-1}(|G|)\cdot q_{\partial(e)-1}(|G|)=r_{\partial(e)}(|G|).
\end{eqnarray*}
As for the second inequality, we claim that any copy of $e$ under $U$ (where $U$ is canonical
for $e$) is an exponential signature containing at most $|G|r_{\partial(e)-1}(|G|)$ instances
of $\rss$ and $\lss$ constructors. To prove that, we can proceed in the usual way (see
the base case above). Now observe that:
\begin{eqnarray*}
W_G&=&\sum_{e\in B_G}\sum_{U\in L_G(e)}(R_G(e,U)-1)\leq\sum_{e\in B_G}\sum_{U\in L_G(e)}q_{\partial(e)}(|G|)\\
   &\leq& \sum_{e\in B_G}\sum_{U\in L_G(e)}q_{\partial(G)}(|G|)\leq\sum_{e\in B_G} r_{\partial(e)}(|G|)\cdot q_{\partial(G)}(|G|)\\
   &\leq& \sum_{e\in B_G} r_{\partial(G)}(|G|)\cdot q_{\partial(G)}(|G|)\leq |G|\cdot r_{\partial(G)}(|G|)\cdot q_{\partial(G)}(|G|).
\end{eqnarray*}
As a consequence, putting $p_n(x)=x\cdot r_n(x)\cdot q_n(x)$ suffices.
\end{proof}
By Proposition~\ref{prop:ELLsound} and Theorem~\ref{theo:first},
normalization of \ELL\ proof-nets can be done in elementary
time, provided $\partial(G)$ is fixed. To this respect,
observe that ordinary encodings of data structures
such as natural numbers, binary lists or trees have bounded
box-depth.
\subsection{Soft Linear Logic}
Soft linear logic (\SLL,~\cite{Lafont04tcs}) can be
defined from \ELL\ by replacing rule $X$ with $M$
as follows:
$$
\infer[M]
{\Gamma,!A\vdash B}
{\Gamma,A,\ldots,A\vdash B}
$$
In proof-nets for \SLL, there are vertices labelled with $M$
and equipped with an arbitrary number of outgoing edges:
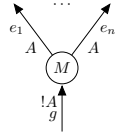
\begin{center}
  \scalebox{0.6}{\epsfbox{figure.81}}
\end{center}
Exponential signatures becomes simpler:
$$
t::=\ess\spb\mss(i)
$$
where $i$ ranges over natural numbers. The new vertex induce
the following rewriting rules:
\begin{eqnarray*}
(h,U,V\cdot \mss(i),+)&\longmapsto_G&(e_i,U,V,+)\\
(e_i,U,V,-)&\longmapsto_G&(h,U,V\cdot\mss(i),-)
\end{eqnarray*}
It can be easily verified that for every $e\in B_G$ and for every
$U\in\mathscr{E}^*$, it holds that $R_G(e,U)\leq |G|$.
Indeed, if $(e,U,t\cdot V,b)\longmapsto_G^*(g,W,Z,c)$,
then $Z=t\cdot Y$. As a consequence:
\begin{proposition}[\SLL\ Soundness]\label{prop:SLLsound}
For every $n\in\N$ there is a polynomial 
$p_n:\N\rightarrow\N$ such that
$W_G\leq p_{\partial(G)}(\length{G})$ for every \emph{\SLL} 
proof-net $G$.
\end{proposition}
\begin{proof}
Simply observe that $R_G(e,U)\leq |G|$ and $|L_G(e)|\leq |G|^{\partial(e)}$. 
As a consequence:
\begin{eqnarray*}
W_G&=&\sum_{e\in B_G}\sum_{U\in L_G(e)}(R_G(e,U)-1)\leq\sum_{e\in B_G}\sum_{U\in L_G(e)}|G|\\
  &\leq&\sum_{e\in B_G}|G|^{\partial(e)+1}\leq \sum_{e\in B_G}|G|^{\partial(G)+1}\leq |G|^{\partial(G)+2}.
\end{eqnarray*}
But this impiles $p_n(x)$ is just $x^{n+2}$.
\end{proof}
\subsection{Light Linear Logic}
Light linear logic (\LLL,~\cite{Girard98ic}) can be
obtained from \ELL\ by enriching the language of formulae
with a new modal operator $\S$ and splitting rule
$P_!$ into two rules:
$$
\begin{array}{ccc}
\infer[S_!]
{!\Gamma\vdash !B}
{\Gamma\vdash B & |\Gamma|\leq 1}
& &
\infer[S_\S]
{!\Gamma,\S\Delta\vdash \S A}
{\Gamma,\Delta\vdash A}
\end{array}
$$
At the level of proof-nets, two box constructions,
$!$-boxes and $\S$-boxes, correspond to $S_!$ and $S_\S$. 
As for the underlying context semantics,
$!$-boxes induce the usual rewriting rules on $C_G$
(see Table~\ref{table:rwexpo}), while the last rule and
its dual are not valid for $\S$-boxes. This enforces \emph{strong
determinacy}, which does not hold for \MELL\ or
\ELL: for every $C\in C_G$, there is at most one context
$D\in C_G$ such that $C\longmapsto_G D$. As a consequence,
weights can be bounded by appropriate polynomials:
\begin{proposition}[\LLL\ soundness]
For every $n\in\N$ there is an polynomial 
$p_n:\N\rightarrow\N$ such that
$W_G\leq p_{\partial(G)}(\length{G})$ for every \emph{\LLL} 
proof-net $G$.
\end{proposition}
\begin{proof}
Observe that, by the subterm property and by stratification, to every 
copy of $e\in B_G$ under $U\in\mathscr{E}^*$ (where $U$ is canonical
for $e$) it corresponds $g\in E_G$ and $V\in\mathscr{E}^*$ such that
$\length{V}=\length{U}$ and 
$$
(e,U,t,+)\longmapsto_G^*(g,V,\ess,+).
$$
Contrarily to \ELL, this correspondence is injective. If, by way
of contraddiction,
\begin{eqnarray*}
(e,U,t,+)\longmapsto_G^*(g,V,\ess,+)\\
(e,U,u,+)\longmapsto_G^*(g,V,\ess,+)
\end{eqnarray*}
where $t\neq u$, then, by duality
\begin{eqnarray*}
(g,V,\ess,-)\longmapsto_G^*(e,U,t,-)\\
(g,V,\ess,-)\longmapsto_G^*(e,U,u,-)
\end{eqnarray*}
But remember that now we have strong determinacy; this
implies either $(e,U,t,+)\longmapsto_G^*(e,U,u,+)$
or $(e,U,u,+)\longmapsto_G^*(e,U,t,+)$. This cannot
be, because of acyclicity. We can now proceed
exactly as in Proposition~\ref{prop:ELLsound}.
Functions $r_n,q_n:\N\rightarrow\N$ are the following ones:
\begin{eqnarray*}
\forall x.r_0(x)&=&1;\\
\forall n.\forall x.q_n(x)&=&|G|\cdot r_n(x);\\
\forall n.\forall x.r_{n+1}(x)&=&r_n(x)\cdot q_n(x).
\end{eqnarray*}
The inequalities:
\begin{eqnarray*}
|L_G(e)|&\leq& r_{\partial(e)}(|G|);\\
R_G(e,U)&\leq& q_{\partial(e)}(|G|);
\end{eqnarray*}
can be proved with the same technique used in Proposition~\ref{prop:ELLsound}.
Letting $p_n(x)=x r_n(x) q_n(x)$ concludes the proof.
\end{proof}

\section{Conclusions}
In this paper, we define a context semantics for linear logic
proof-nets, showing it gives precise quantitative information
on the dynamics of normalization. Theorems~\ref{theo:first} and
\ref{theo:second} are the main achievements of this work: they 
show that the weight $W_G$ of a proof-net $G$ is a \emph{tight}
estimate of the time needed to normalize $G$. Interestingly,
proving bounds on $W_G$ is in general easier than bounding normalization
time by purely syntactic arguments. Section~\ref{sect:subsystems}
presents some evidence supporting this claim.

Results described in this paper can be transferred to affine logical
systems, which offer some advantages over their linear
counterparts (for example, additive connectives can be expressed in the logic).

An interesting problem (which we leave for future work) is characterizing 
the expressive power of other fragments of \MELL, such as \QLL\ or \TLL\ (see~\cite{Danos03ic}),
about which very few results are actually known. We believe that the
semantic techniques described here could help dealing with them.
Any sharp result would definitely help completing the picture.

Interestingly, the way bounded linear logic (\BLL, \cite{Girard92tcs}) is defined is very 
reminiscent to the way context semantics is used here. We are currently
investigating relations between the two frameworks.

\subsection*{Acknowledgments}
The author wishes to thank Patrick Baillot and Olivier Laurent for many interesting 
discussions about the topics of this paper

\bibliographystyle{latex8}

\end{document}